# Evaluating the Environmental Justice Dimensions of Odor in Denver, Colorado

Priyanka N. deSouza[1,2*], Amanda Rees[1], Emilia Oscilowicz[1], Brendan Lawlor[3], William Obermann[3], Katherine Dickinson[4], Lisa M. McKenzie[4], Sheryl Magzamen[5,6], Shelly Miller[7], Michelle L. Bell[8]

1: Department of Urban and Regional Planning, University of Colorado Denver, Denver CO, 80202, USA
2: CU Population Center, University of Colorado Boulder, Boulder, CO 80302, USA
3: Denver Department of Public Health and Environment, Denver CO, 80202, USA
4: Department of Environmental and Occupational Health, Colorado School of Public Health, University of Colorado Anschutz, Aurora, CO, 80045, USA
5: Department of Environmental and Radiological Health Sciences, Colorado State University, Fort Collins, CO, USA
6: Department of Epidemiology, Colorado School of Public Health, Colorado State University, Fort Collins, CO, USA
7: Dept. of Mechanical Engineering and Program of Environmental Engineering, Univ. of Colorado Boulder, Boulder, CO 80309
8: School of the Environment, Yale University, New Haven, CT, 06520-8354, USA

*: priyanka.desouza@ucdenver.edu

## Abstract

**Background**
Odors are a documented environmental justice challenge in Denver County, Colorado, USA. Complaints are an important modality through which residents express their concerns.

**Objective**
We investigated disparities in environmental justice related-variables based on home *and* workplace census block groups (race/ethnicity, education-levels, renter-occupied housing, median income and median home values, gentrification) based on locations of odor complaints as well as that of potentially malodorous facilities. We report key themes identified in complaints.

**Methods**
We obtained odor complaints for 2014-2023, and the locations of facilities required to submit odor management plans as of 2023 from the Denver Department of Public Health and Environment. We downloaded residential census block group-level socioeconomic data from the 2016-2020 American Community Survey; and workplace based socioeconomic data from the Longitudinal Employer-Household Dynamics dataset for 2020. We categorized neighborhoods as gentrified or not based on a typology produced by the Urban Displacement Project. We assessed exposure to potentially malodorous facilities and complaints within each census block group and investigated exposure disparities by comparing distributions of environmental justice-related variables based on if a facility or a complaint has been made, and census block group-level odor intensity categories. We used unsupervised machine learning to identify themes from the odor complaints.

**Results**
Less privileged census block groups were significantly disproportionately burdened with potentially malodorous facilities. Importantly, our study also reveals disparities in the location of facilities, not just traditional residence/-based environmental justice related variables, but in workplace/-based factors, as well. We did not observe similar disparities for odor complaints. However, complaints were significantly higher in gentrified neighborhoods. Specific facilities were mentioned repeatedly in the complaints received.

**Significance**
Our study adds to the growing literature on disparities associated with odor and odor complaints, and suggests the need for further research using detailed exposure assessment in other locations.

# 1 Introduction

Air pollution is one of the world's biggest environmental health risks[1–3] and is responsible for nearly seven million premature deaths worldwide annually[4]. Ambient odors, a type of air pollution that has been linked with negative effects on health and wellbeing[5]. Exposure to odor has been shown to be associated with symptoms such as headaches, nausea, fatigue, eye and throat irritation, shortness of breath, inability to concentrate and stress[6]. More research is needed to better characterize the health impacts of odor and potential disparities.

Ambient odors are also an important social issue and are the source of many air quality-related complaints. Odor-emitting facilities are often located adjacent to communities of minority, lower-income, working class, and socioeconomically marginalized residents[7–10]. In the United States, strategic urban planning policy, zoning, and redlining have contributed to these siting patterns[11,12]. Thus, researchers have argued that odor pollution must be researched and addressed through a lens of environmental justice, considering socioeconomic vulnerabilities including race, income, and gender[13,14].

Regulatory efforts at the municipal, state, and federal level have been enacted to address odors and their negative social and health impacts [15]. In the United States, most of these regulatory actions were performed at the municipal level[16]. A recent approach at the municipal level to identify malodors is the collection of general complaints from affected individuals through a self-reporting system via an online portal or through a telephone hotline[17]. These methods, while inexpensive and relatively easy to implement, suffer from risk of individualized bias, where one person may perceive an odor to be more offensive than another person, as well as difficulty in verification of exact geolocation, time, and other critical identifying factors[17]. While self-reporting of odor complaints has been identified as an important modality to address odor, socioeconomic biases in the reporting of complaints have largely not been studied, despite the importance of these considerations for municipalities and urban planners when engaging in reparative, equitable, inclusive, and participatory planning practice[18].

In the northern neighborhoods of Denver, Colorado, odor is a documented environmental justice concern that disproportionately impacts lower-income and minority residents[19–21]. Residents in these neighborhoods have identified odor as a major detriment to quality of life, directly impacting personal health[20]. Odor polluting industries in the northern Denver neighborhoods remain under intense scrutiny by the community and environmental justice advocates[22,23]. In this study, we aim to investigate the associations between the distribution of odor complaints over time, as well as that of potentially malodorous facilities in relation to environmental justice related variables in Denver county. This is to our knowledge, is the first study on odor that has considered the social and environmental justice implications of self-reporting of odor complaints, *and* one of the first to do so with the locations of potentially malodorous facilities. Recent research has demonstrated important disparities in air pollution over both home and workplaces[24]. We intend to extend this burgeoning thread of research to see if we see similar disparities in exposure to ambient odor at workplaces. By reporting disparities in workplace *and* home-based locations, we aim to explore if in addition to housing segregation and other residence-based discrimination patterns such as industrial facility siting, other workplace-related forces of structural racism are also in play in generating inequalities in the US. Workplace related forces of structural racism contributing to these disparities might include disparities in access to transportation infrastructure and/or limited or constrained range of accessible workplaces. Finally, we also evaluated key themes that arose in user complaints to better assess the range of issues related to odor across Denver.

# 2 Data and Methods

We evaluated the disparities in the distribution of potentially malodorous facilities (2023) and odor complaints (2014-2023) in Denver County at the block group level by different environmental justice related variables: racial and ethnic groups (White, Black, Hispanic for residents as well as workers), education levels (less than high school for residents and workers), and owner/renter-occupied housing; and median household income and median home values. We also report key themes in the odor complaints received.

## 2.1 Potentially Malodorous Facilities

We obtained the addresses of 269 potentially malodorous facilities required to submit an Odor Control Plan from the Denver Department of Public Health and Environment (DDPHE) in 2023 (**Figure 1**). As per the 2016 updated Denver air pollution control ordinance[25], facilities that are required to submit Odor Control Plans fall in the following categories: 1) Receives five or more complaints from separate households or businesses within Denver County in a 30-day period;

2) Exceeds the one-to-seven dilution threshold; 3) Engages in activities such as pet food manufacturing, marijuana growing, processing, or manufacturing, meat byproduct processing, asphalt manufacturing, petroleum refining, sewage treatment, wood preservation; or 4) was asked to submit a plan by the Manager of the Denver's Department of Environmental Health (https://www.denvergov.org/files/assets/public/v/1/public-health-and-environment/documents/eq/odor-faq-final.pdf, Last accessed 8 Dec, 2023). As the focus of our analysis is on Denver County, we excluded three facilities located in Arapahoe County and one located in Adams County, which were out of our study area, leaving us with 265 facilities. Facilities were categorized as marijuana- (257) and non-marijuana- (8) related. All addresses were geocoded by the authors using R 4.3.1.

## 2.2 Odor Complaints

We obtained data on 1,322 odor complaints received between the years Jan 2014 - May 2023 by DDPHE. We excluded the 33 complaints made in Adams County and the six in Arapahoe County, as they were outside our study area, leaving us with 1,283 complaints. For all but five complaints (three complaints in 2023, one complaint in 2022 and one in 2015) we had information on the date the complaint was received and the data the investigation was 'closed'. We generated a new variable: time (days) between the date each complaint was finalized or 'closed' and the date each was received. For each complaint we also had information on the address and complaint text, where all personal/identifiable information was removed by DDPHE staff.

Complaints were logged via Denver's Online Services Hub (formally Pocketgov) or 311 calls (non-emergency calls). All locations of complaints that residents believed were the cause of the odors reported, or residents' best guesses, were geocoded by the authors using R 4.3.1. In our main analysis, we considered the sum of all complaints made in this time period (**Figure 1**). In supplementary analyses, we evaluated the distribution of complaints recorded each year separately (**Table S1**).

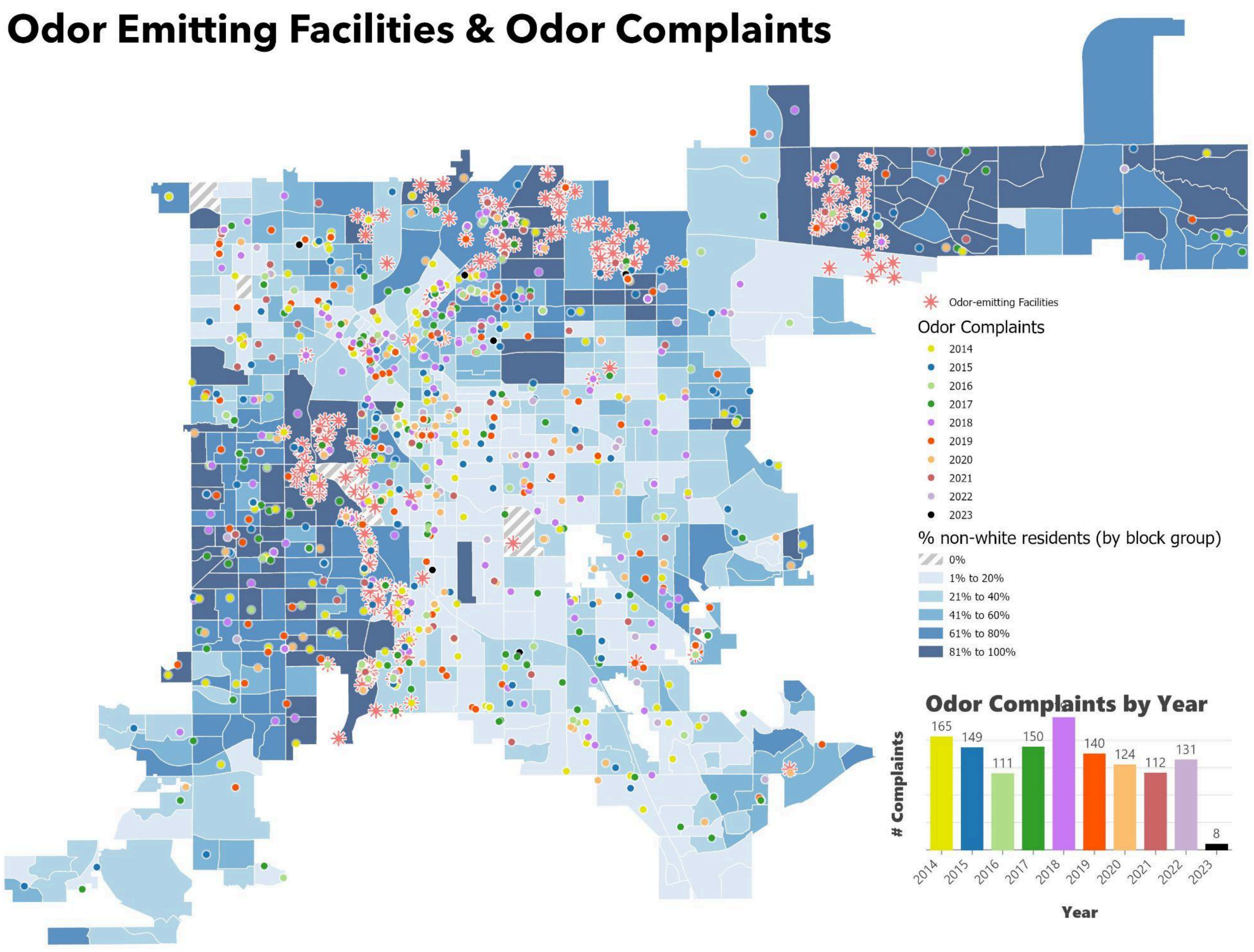

**Figure 1:** *Locations of odor complaints and potentially malodorous facilities in Denver.*

## 2.3 Environmental justice related variables

We selected various environmental justice related variables that are commonly used in previous literature[26,27], although we recognize that numerous other variables reflecting different concepts also exist (e.g., racial residential segregation)[28]. In addition to using residence-based environmental justice related variables, we also used workplace-related variables to evaluate if disparities in exposure to odors were also experienced by workers. We acknowledge that we cannot distinguish if the 311 odor complaints were filed by residents or workers. However, the location of the odor complaints serves as a proxy of important sources of odor that can impact health and wellbeing. We thus evaluate if disparities exist in resident and worker exposure to odors.

### 2.3.1 Residential

We downloaded socioeconomic data (SES) for each census block group (using block groups defined in the 2020 decennial census) for Denver from the American Community Survey (ACS) 2016-2020. Specifically, we downloaded the proportion of residents by racial and ethnic groups (White non-Hispanic, Black non-Hispanic, Hispanic), percent of residents with less than a high-school edcuation, high school, percent renter-occupied housing; median household income and median home value.

### Gentrification

DDPHE inspectors have noticed an uptick in odor complaints in rapidly gentrifying neighborhoods in Denver such as RiNo[29]. They posited that this was likely the case because the newer residents who are not accustomed to odors might not respond in the same way as older residents. Gentrification, the process of change and displacement of a lower wealth population by a higher wealth population, is widely deemed as an important environmental justice issue[30]. We leveraged a widely-used indicator of gentrification developed at the census tract-level for the time period 2000-2017 by the Urban Displacement project (UDP) [31] that categorizes census tracts into: a) Low-income/Susceptible to displacement, b) Ongoing displacement, c) At risk of gentrification, d) Early/ongoing gentrification, e) Advanced gentrification, f) Stable moderate/mixed income, g) At risk of becoming exclusive, h) Becoming exclusive, i) Stable/advanced exclusive, j) High student population, and j) Unavailable/unreliable data. UDP defines low-income neighborhoods as those with a median household income below 80% of the regional mean and moderate-to-high-income neighborhoods as those at or above 80% of the regional median income. UDP classified neighborhoods as exclusionary if rents were so expensive that low-income residents were excluded from moving in.

We simplified the UDP categories in Gentrification to three categories: Early/ongoing gentrification, Advanced gentrification and Not Gentrification (all other categories). The UDP data are available at the 2010 decennial census tracts, which we used for gentrification-related analyses. Due to differences between the 2010 and 2020 decennial census tracts, the total number of complaints across 2014-2023 within the 2010 decennial census tracts boundary in Denver County were 1,261. Of these, 21 complaints are in census tracts with unreliable/unavailable data about gentrification and were excluded from our main analysis. Because of the different spatial scale of the gentrification indicator, we analyzed the impact of gentrification on odor complaints, separately from the other residence-based environmental-justice variables considered.

## 2.3.2 Work-related

We downloaded Workplace area characteristics (WAC) at the census block level from the Longitudinal Employer-Household Dynamics (LEHD) Origin-Destination Employment Statistics (LODES) version 8 data product for Denver for the year 2020. WAC data contain the total number of jobs (primary and secondary) for each census block, as well as jobs disaggregated by environmental justice related categories such as, race (White, Black), ethnicity (Hispanic), formal educational attainment (less than high school), among others. We aggregated the LODES WAC data to the census block group level for Denver. Note that racial/ethnicity categories in the LODES dataset are different from the ACS data; White and Black Americans in the former include Hispanic Americans while the latter provides information on White, non-Hispanic, and Black non-Hispanic Americans. Also note that the LODES population comprises only workers, while the ACS data product comprises all Americans residing in a given census block group.

## 2.4 Methods

### 2.4.1 Time Variation of Complaints

We describe the distribution of complaints over time by reporting the average number of complaints received between Jan 2014 - May 2023.

### 2.4.2 Spatiotemporal Clustering of Complaints

DDPHE inspectors argued that the rise in social media also likely impacts the distribution of complaints. Specifically, they posited that when people share information with their neighbors or coworkers on how to make a complaint, more odor complaints are received from specific locations. We evaluated spatiotemporal clustering of odor complaints using the spotoroo package in R[32]. The algorithm splits the data into fixed time intervals, according to a user defined time step, which we set as one day. Within each time window, hotspots that fall within defined spatial thresholds (maximum intra-cluster spatial distance between nearest hotspots was set to 1 km, and the minimum number of hotspots in a cluster was set to five), were identified using hierarchical clustering. We plot the timeline of hotspots identified over the course of this study.

### 2.4.3 Identifying Topics

We use topic modeling, an unsupervised machine learning technique[33–35], to extract key topics from the 1,322 odor complaints received. Unsupervised machine learning has been commonly used in studies that do not have any predetermined framework to analyze unstructured text data[33–35]. As this process allows topics to be identified from the data, it may avoid biases generated through non-automated coding that relies on subjective interpretations of the text. Briefly, topic models pick up co-occurrence signals between different words in the collection of texts. The underlying assumption is that words that occur often in the same sentence are likely to belong to the same latent topic. A topic model represents the overall themes or topics in a corpus as probability distributions over words in a vocabulary; so, while the probability of the word “dust” is high in a topic relating to construction, it will likely be relatively low in a topic pertaining to marijuana. Documents (which in this case are complaints) consisting of combinations of words were modeled using a generative process where first a topic is selected according to some probability distribution specific to a given document, and then, a word is selected from that topic in accordance with the topic’s distribution over vocabulary words. Using the documents in this corpus (which are the output of such a generative process), we can infer the likelihood of each topic given a document and each word given a topic through a training process.

We preprocessed the data by removing special characters (non-ASCII characters), punctuation, and numbers were also removed from the review text. Stop words, such as ‘the’ and ‘of’ from the SMART stopwords list, which are built into the tidytext package in R, were removed. After this step, additional stop words from a custom list were removed: {"odor", "say", "like", "yes", "say", "caller", "call", "online", "case", "complaint", "complainant's", "Complainant's", "Complaint", "redact", "redacted", "answer", "question", "addit", "method", "contact", "email", "descript", "issu", "issue", "emails", "complain", "describ", "come", "online", "onlin", "report", "take", "prefer", "back", "can", "thank", "phone", "address" }. Stop words tend to be high-probability terms that can skew the word-type probability distribution and slow inference from the topic models.

Stemming was used to lemmatize words and their derivatives (e.g., determine, determined, determining), thus rendering all derivative forms of a single word in an unambiguous, non-inflected state. As language exhibits a rich inflectional morphology, if derivatives of the same word are treated as unique tokens, the co-occurrence signal between different words in the corpus under consideration will become weaker. We thus applied stemming to improve the quality of the topic model.

Latent Dirichlet Allocation (LDA) is the most commonly used topic model. However, LDA has several limitations, one of the most important being that it assumes that topics are independent of each other[36]. The recently introduced structural topic modeling (STM) algorithm builds on the LDA model and allows topics to be correlated[37–39]. We implemented the topic modeling in R with the STM package, using the spectral algorithm without the inclusion of covariates[38]. When the number of documents is large, as in this case, the spectral algorithm has been shown to perform well and provide more stable and consistent results than the LDA model[38].

To determine the number of topics in the text, this paper used the metrics of held-out likelihood and semantic coherence provided by the STM package. Held-out likelihood (an indication of cross-validation) is calculated by holding out 10% of the words in the corpus, training the model, and using the document-level latent variables to evaluate the probability of the held-out words. Semantic coherence (an indication of higher topic interpretability) measures the frequency of the co-occurrence of the top words of a topic[40]. This paper compared the performance of 3,4, 5, 6, 7, 8, 9, 10, and 15 topic models using these metrics (**Figure S1**).

From **Figure S1**, we note that four topics appear to provide a good trade-off between the greatest coherence and the held-out likelihood. A model with four topics also withstood the author’s subjective evaluation of the themes produced. This paper provides brief summaries of each topic derived from the four-topic model by reporting words with the highest occurrence probability corresponding to a specific topic.

### 2.4.4 Environmental Justice Analysis

We merged the complaints and the database of potentially malodorous facilities with the corresponding block group in which each complaint or facility was registered. Ten of the total 571 census block groups (**Figure 1**) had no residents. 22 complaints (two in 2014, six in 2015, one in 2017, three in 2018, three in 2019, one in 2021, and four in 2022) and 42 facilities (41 marijuana and 1 non-marijuana) were located in block groups without residents. When evaluating disparities in environmental justice related variables based on residential characteristics, we excluded these census block groups and the corresponding complaints and facilities from our analysis. Four of the total 571 census block groups had no workers in Denver.

One complaint in 2014 was located on one of these block groups. All facilities were located in block groups with workers present. When evaluating disparities in environmental justice related variables using workers based characteristics, we excluded these census block groups and the corresponding complaint from our analysis.

Pearson correlation coefficients were used to evaluate correlations between all block group environmental justice related variables based on residential and workplace locations for all of Denver county, separately. We mapped the spatial distribution of odor complaints and potentially malodorous facilities and environmental justice related variables to examine their spatial relationships.

We assessed exposure to odor at the block group level using the following metrics:

(1) Absence/Presence:
  (a) If any complaints were registered in a given block group
  (b) If any facility was located in a given block group
(2) Per person:
  (a) Total complaints per resident (worker) in each block group
  (b) Facilities per resident (worker) in each block group
(3) Per area:
  (a) Total complaints per $km^2$ in each block group
  (b) Total facilities per $km^2$ in each block group
(4) Mean number of days taken for a complaint to be resolved in each block group.

These metrics allow us to consider the density of odor complaints, facility siting, and resolution time in each block group. For metric (1), we compared the distribution of environmental justice-variables in block groups with registered complaints compared to others, and separately in block groups with potentially malodorous facilities to those without, using Mann-Whitney U test (all of the environmental justice related variables considered were not normally distributed for our study area as determined by a Shapiro-Wilk test).

For metrics, (2), (3), and (4), we categorized block groups into four categories based on the continuous exposure metrics from low exposure (quartile 1) to high exposure (quartile 4). **Table S2** provides the range of each exposure corresponding to each quartile of each metric. We investigated exposure disparities to odors by comparing the distribution of environmental justice related variables based on the exposure intensive categories (i.e., from no exposure and low exposure: quartile 1 to high exposure: quartile 4). We compared differences in the distributions by exposure categories (quartile 1 to quartile 4) using Kruskal-Wallis tests (all of the environmental justice related variables considered were not normally distributed for our study area as determined by a Shapiro-Wilk test). For multiple pairwise comparisons for continuous variables, we performed post hoc tests using Dunn tests. We used Jonckheere-Terpstra tests to test trends by quartiles of exposure intensity.

Note, we also conducted our analysis of the distribution of facilities in Denver disaggregated by facility type (marijuana: 261 facilities or non-marjuana related: 8 facilities) to evaluate if facility-type may affect these findings. As there were only 8 non-marijuana facilities, we used Mann-Whitney tests to evaluate if environmental justice related variables in block groups with these facilities differed significantly from those without.

In order to evaluate the impact of gentrification on the distribution of odor complaints, we aggregated complaints received to the census-tract level. We then used the Mann-Whitney U test to report if there were significant differences in the distribution of the number of complaints as well as the density/intensity of complaints (complaints/$km^2$ and complaints/population) by the simplified UDP gentrification typology (gentrification/not gentrification).

Statistical significance in this study was assessed using a $p < 0.05$ threshold. All analyses were conducted using the software: R 4.3.1. All maps were created using QGIS 3.32.3.

# 3 Results

## 3.1 Time Variation of Complaints

The average diurnal, day-of-the-week, and monthly variation in complaints for each year are displayed in **Figures S2-S11**. Large variations in the temporal patterns of the complaints are observed every year. For example, for the year 2014, no complaints were recorded over the weekend, while for the year 2015, no complaints were recorded on Friday or Saturday. For every following year, complaints were recorded on the weekend. Part of the reason for the differences in the variation of complaints over a week was the difference in managing odor complaints in 2014-15 compared to later years. In 2014-15, complaints received from 311 calls were manually entered by DDPHE staff, which meant that they were recorded on weekdays during normal working hours. Since 2015, Salesforce software for tracking odor concerns was used instead of 311 and DDPHE staff logging. The Salesforce system records the complaints automatically as soon as they are received. This allows the recording to happen at any time, even on weekends or outside of normal working hours.

## 3.2 Spatiotemporal clustering of Complaints

13 hotspots of odor complaints (at least five complaints observed within the same 1 km radius over the course of a day) between Jan 2014 - May 2023 were observed (**Figure 2**). **Figure 2** also indicates that the number of hotspots have steadily increased over time, with the bulk of the hotspots observed post 2018.

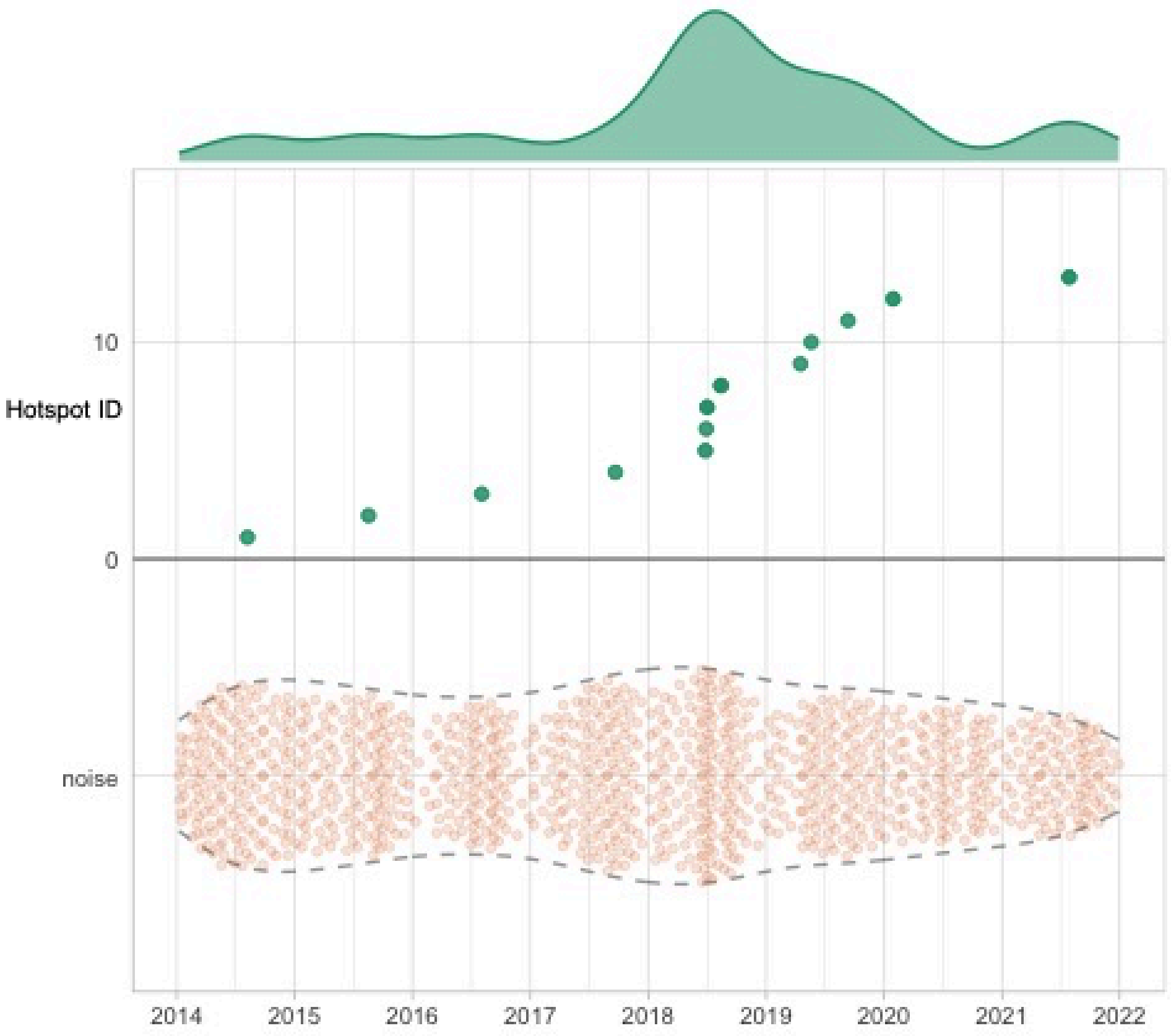

***Figure 2:*** *Timeline plot for providing an overview of the hotspots of odor complaints identified over 2014-2023. The x-axis represents the date and the y-axis is the cluster membership.The observed time of hotspots are shown as dot plots (green). The density plot at the top displays the temporal frequency of complaint occurrence over the time period considered, The dot plot at the bottom (orange) shows the number of complaints that are considered to be noise as they do not comprise a hotspot.*

## 3.3 Topics Identified in the Complaints

The word clouds for each topic depicting the words with the highest probability of occurrence are presented in **Figure 3**. The larger the font, the higher the probability of the occurrence of the word for the given topic. Briefly, it appears that the most prevalent topics discussed in complaints related to the pet-food manufacturing factory Purina, neighborhood smells, construction and building-related smoke, and marijuana-related odor complaints.

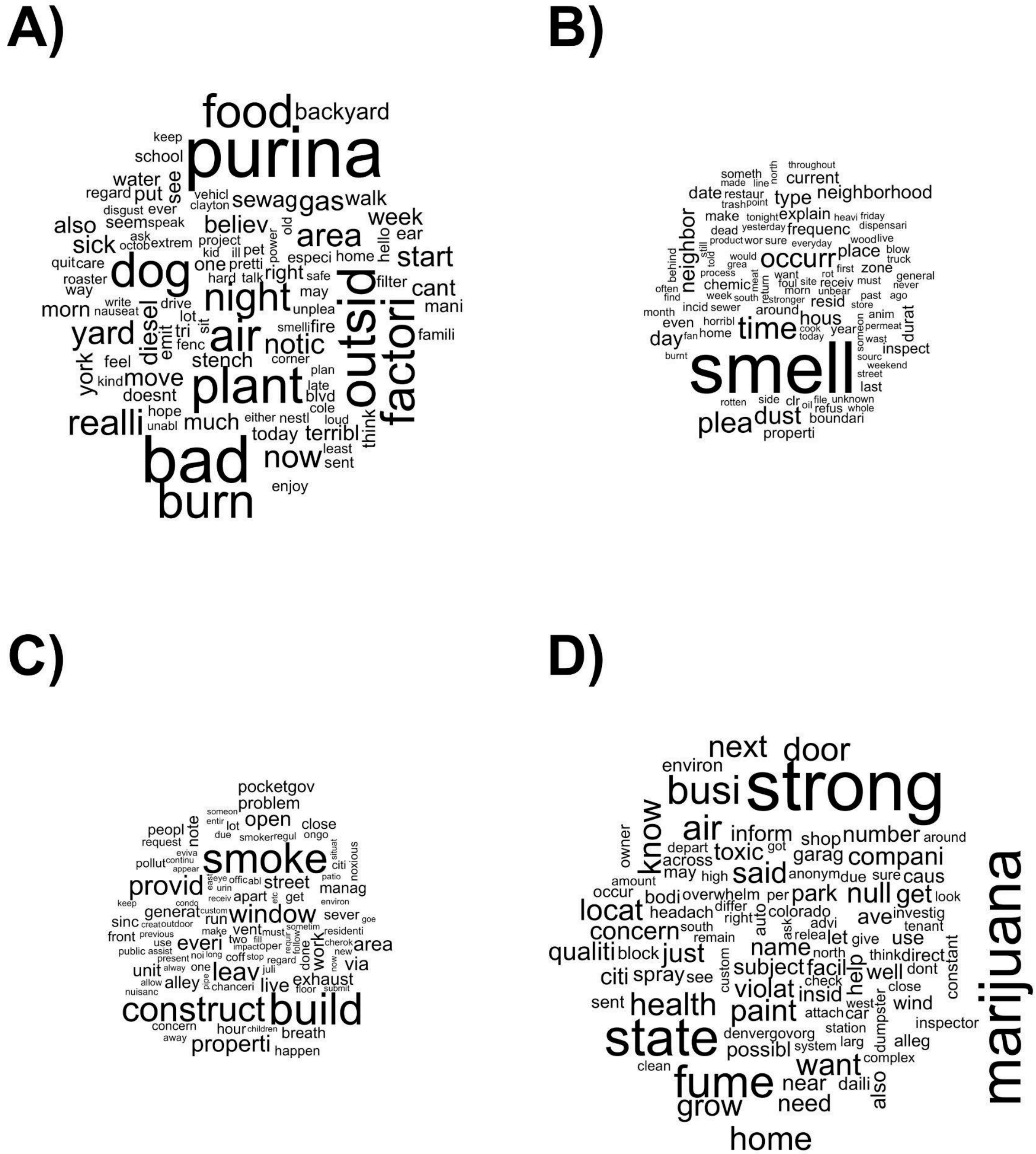

*Figure 3*: *Word clouds for each topic: A) Smells from Purina, a factory that produces dog and cat food, B) Smells reported from neighborhood sources, C) Smoke-related concerns from construction and other work in the area, D) Concerns related to marijuana and other smaller industrial sources such as fumes from paint, etc. Words with the highest probability of occurrence in each topic are shown. Words in a larger font size indicate a higher occurrence likelihood. Topics are manually labeled by the authors. Note that stemming was performed to*

*ensure that all conjugations of a word were reduced to a single entity. For example, environments, environment, and environmental were reduced to environ.*

## 3.3 Environmental-Justice Analysis

The distribution of odor-emitting facilities highlighted in **Figure 1** demonstrates the “inverted-L pattern” of geographic disparities in Denver. These areas, once home to railyards, smelters, and stockyards, are separated from the rest of the city by two major Interstate highways — Interstate 25 runs north-south along the Platte River Valley and Interstate 70 runs east-west in the northern part of the city (note that I-70 running through North Denver just completed a 5-year renovation, which may explain an increase in construction odors). They are also adjacent to or surrounded by land used for industrial purposes, including marijuana growhouses. Socioeconomically, these neighborhoods have fewer years of formal education, earn a lower median income, and have higher proportions of non-White residents. **Figure 1** also displays the locations of all odor complaints from 2014 to 2023. Although the points are distributed throughout the city, many of them are located within the inverted-L.

**Figure S12- S16** displays maps of the environmental justice related variables used in this study. Pearson pairwise correlation coefficients between the environmental justice related variables at the census block group level for Denver are displayed in **Figure S17** for residents and workers, respectively.

For residence-based environmental justice related variables, the highest correlations were observed between the percentage of Hispanic residents and the percentage of White residents (R: -0.77), between the percentage of Hispanic residents and the percentage of residents with formal education < high school (R: 0.76), between the percentage of White residents and percentage of residents with formal education < high school (R: -0.72). Other pairwise correlations fell in a wide range, from weak correlations between the percentage of Black residents and the percentage of renter occupied homes (R: 0.05) to moderate correlations, such as between median household income and percentage of renter occupied homes (R: -0.49), and percentage White residents and median home value (R: 0.43), for example (**Figure S17**). For work-based environmental justice related variables, the Pearson correlation coefficients were high between the percentage of White workers and Black workers (R: -0.76), percentage of Hispanic workers and workers with formal education < high school was 0.66. Other pairwise correlation coefficients were lower with magnitudes < 0.15 (**Figure S17**). Correlations between resident and work place based environmental justice related variables varied considerably. The correlation between % White residents and White workers was 0.43, between % Black residents and Black workers was -0.10, between % Hispanic residents and workers was 0.04, between % residents and workers with formal education < high school was -0.14 (**Figure S17**).

We compared the distribution of environmental justice related variables of block groups that had a facility or registered complaint with those that did not (**Figure 4**). Mann-Whitney U-tests revealed that there were significant differences in block groups without facilities compared to those with facilities in the distribution of the percentage of White residents (without:

mean=59.2%, with: mean= 47.1%), percentage Hispanic residents (without: 26.0%, with: 36.9%), percentage residents with formal education < high school (without: 6.9%, with: 11.0%), percentage renter-occupied homes (without: 44.8%, with: 57.8%), median household income (without: USD$ 81,634, with: USD$ 58,676), and median home value (without: USD$ 430,196, with: USD$ 346,827) (**Table 1**). Tests revealed similar statistically significant differences in the distribution of all work place based environmental justice related variables except % White workers across block groups with facilities compared to those without. Specifically, the mean percentage of Black, Hispanic and workers with formal education < high school in block groups with facilities were 6.8%, 23.6%, and 12.6%, respectively; while the corresponding means in block groups without facilities were 6.5%, 22.0%, and 12.0%, respectively (**Table 1**).

When we repeated the above analyses to evaluate disparities in facility siting for the 261 marijuana facilities, alone (**Table S3**), we observed similar disparities for marijuana-related facility siting. For non-marijuana related facilities, we observed that the distribution of median annual household income (without: USD$ 80,257, with: USD$ 32,644), was significantly different between block groups that had such facilities versus across those that did not.

No statistically significant differences were observed in the distribution in home and work place based environmental justice related variables between block groups with a complaint registered and those without (**Table 1**). In analyses conducted on an annual-basis between 2014 - 2022, we did not observe any significant differences in the distribution of any environmental justice related variable between block groups with complaints compared to those without (**Figures S18-19**; **Table S4**).

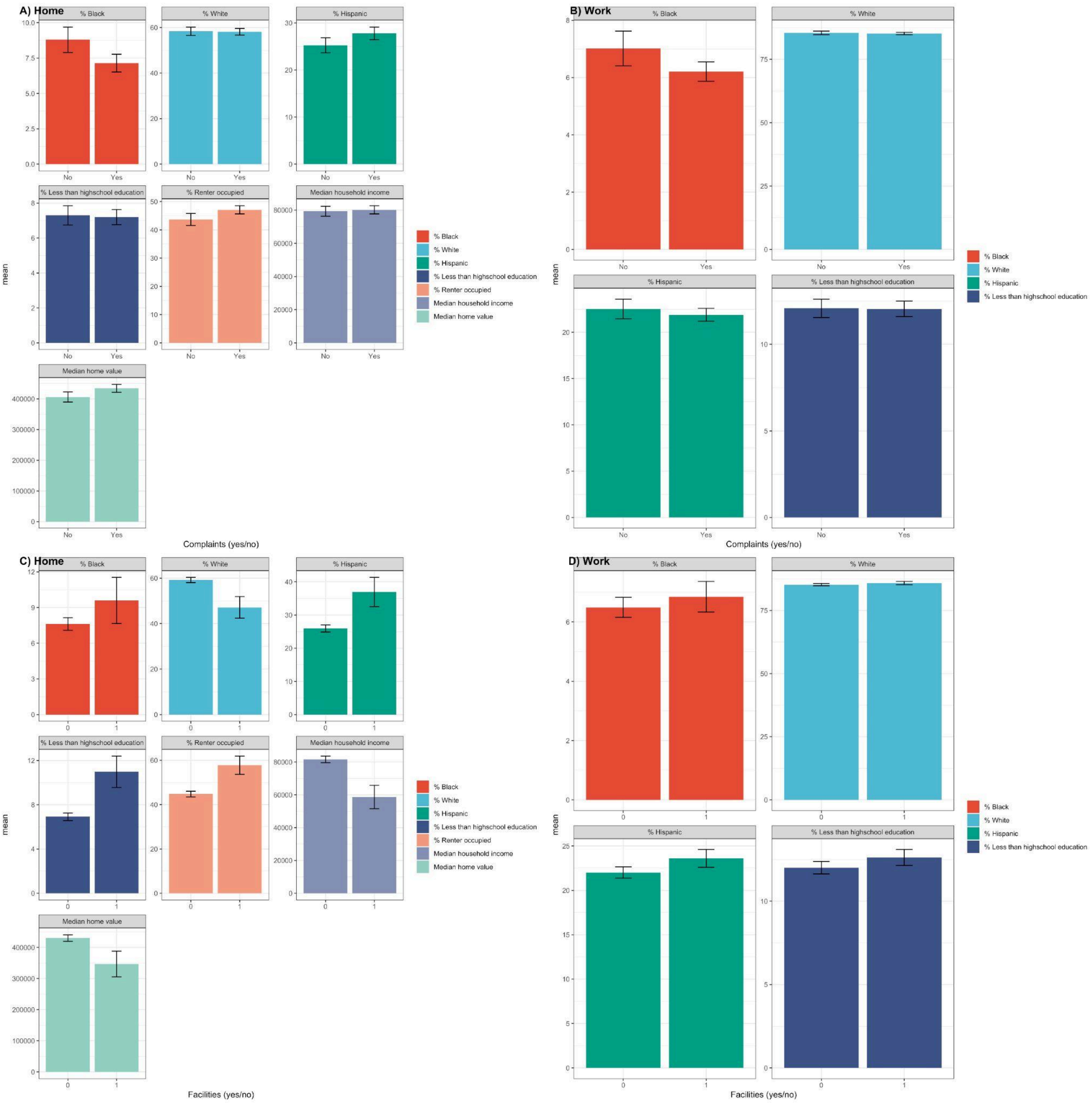

***Figure 4**: Distribution (Mean with SE error bars) of a) residence and b) workplace-based environmental justice related variables for census block groups with complaints registered over the time period Jan 2014 - May 2023, and c) residence and d) workplace-based environmental justice related variables for census block groups with potentially malodorous facilities in 2023 in Denver. Number of census block groups with a facility registered = 45, Number of block groups with a complaint = 351, Total block groups = 571.*

The disparities observed varied when we considered odor exposure intensity/density patterns (**Figure 5**; **Table 1**). Within block groups with facilities, no discernible trend was detected in the environmental justice related variables considered across quartiles corresponding to facilities per population; However, across quartiles corresponding to facilities per km$^2$, we observed significant differences in census block groups corresponding to the percentage of residents with formal education < high school (means: quartile 1: 7.8%, quartile 2: 15.1%, quartile 3: 15.0%, quartile 4: 6.0%) (**Table 1**). Jonchkeere-Terpstra's test revealed no significant trends across block groups corresponding to different quartiles of either facility-based exposure metric (**Table 1**). When we repeated this analysis disaggregated by facility type, we observed similar trends in the distribution of environmental justice related variables for marijuana facilities (97% of all facilities in our dataset) siting (**Table S3**).

Within block groups with registered complaints, we found significant differences in census block groups corresponding to different quartiles based on the number of complaints per $km^2$ for percentage of residents with formal education < high school (means: quartile 1: 7.4%, quartile 2: 7.7%, quartile 3: 8.1%, quartile 4: 5.5%), and percentage of renter occupied homes (means: quartile 1: 36.6%, quartile 2: 43.1%, quartile 3: 50.9%, quartile 4: 57.8%), and percentage Hispanic workers (means: quartile 1: 22.9%, quartile 2: 25.3%, quartile 3: 20.8%, quartile 4: 18.7% ) (**Figure 5**; **Table 1**). The pattern of the distribution of residence and workplace environmental justice related variables was not consistent. Specifically, census block groups with the highest density of complaints had the lowest % of residents with formal education < high school, the lowest number of Hispanic workers, and the highest number of renter occupied homes. Jonchkeere-Terpstra's test reveals trends in the distribution of percentage of renter occupied homes, and median household income, and percentage Hispanic workers across block groups belonging to different quartiles corresponding to the number of complaints per $km^2$ (**Table 1**).

We also observed significant differences in census block groups corresponding to different quartiles based on the number of complaints per population for the percentage of Black workers (mean: quartile 1:6.6%, quartile 2: 5.8%, quartile 3: 6.5%, quartile 4: 5.9%) (**Figure 5**; **Table 1**). Jonchkeere-Terpstra's test reveals that trends were observed in the distribution of percentage Black workers, and percentage of workers with formal education < high school across block groups belonging to different quartiles corresponding to the number of complaints per population (**Table 1**).

In analyses conducted on an annual-basis between 2014 - 2022, we observed different patterns of disparities over time (**Table S5**). Generally, however, we observed significant differences in the distribution of complaints recorded per $km^2$ and complaints per population for the following environmental justice related variables over multiple years: percentage white residents, percentage residents with formal education < high school, percentage renter occupied homes, percentage Black workers, percentage Hispanic workers and percentage workers with formal education < high school (**Figures S20-S23; Table S5**). Here too, no clear discernable pattern was observed in the different distributions of environmental justice related variables.

We did not observe significant differences in environmental justice related variables across block groups belonging to different quartiles corresponding to the mean time to resolution (days) (**Table 1**). We did not have enough variation in the time to resolution of complaints to repeat this analysis on an annual-basis.

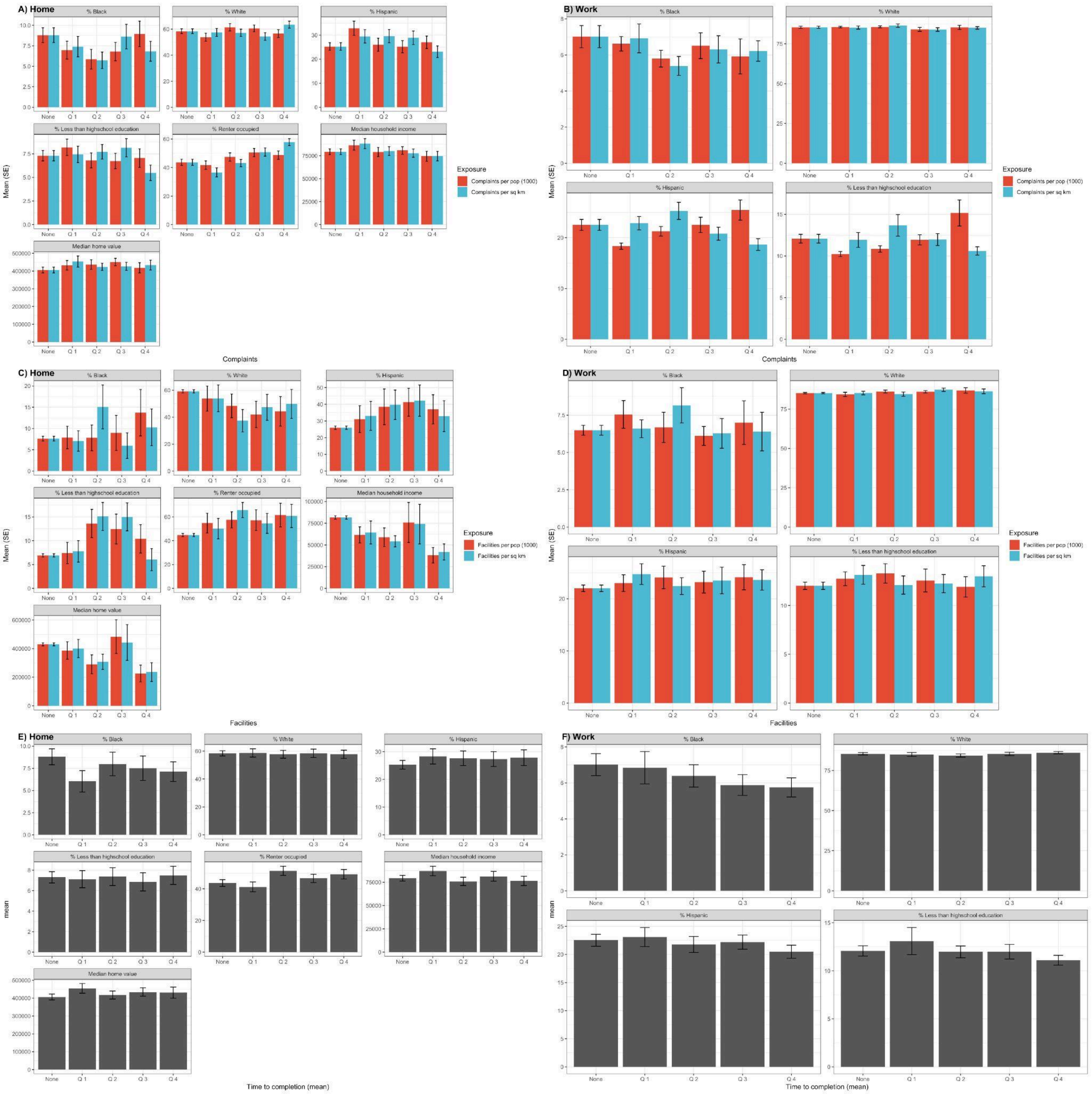

***Figure 5**: Distribution (Mean with SE error bars) of environmental justice related variables over census block groups categorized by quartiles by the different exposure metrics related to A) Total complaints registered between Jan 2014 - May 2023, B) Potentially malodorous facilities and C) Mean time of complaint resolution (days) considered in this study at the census block group level in Denver.*

***Table 1**: Results from statistical tests to compare the distribution of environmental justice related variables by exposure intensity based on different exposure metrics at the census block group level in Denver.*

| | **Complaints** | | | **Time** | **Facilities** | | |
|---|---|---|---|---|---|---|---|
| | **Per km$^2$** | **Per Population** | **Yes/ No** | **Mean time to** | **Per km$^2$** | **Per Population** | **Yes/ No** |

| | | | (1000) | | | completion | | | | (1000) | | |
|---|---|---|---|---|---|---|---|---|---|---|---|---|
| | p-value[a] | p-value trend[b] | p-value[a] | p-value trend[b] | p-value[c] | p-value[a] | p-value trend[b] | p-value[a] | p-value trend[b] | p-value[a] | p-value trend[b] | p-value[c] |
| **Residential** | | | | | | | | | | | | |
| **% Black** | 0.66 | 0.84 | 0.39 | 0.86 | 0.30 | 0.38 | 0.26 | 0.66 | 0.67 | 1.00 | 0.98 | 0.28 |
| **% White** | 0.17 | 0.37 | 0.23 | 0.71 | 0.86 | 0.97 | 0.91 | 0.69 | 0.94 | 0.76 | 0.38 | 0.01 |
| **% Hispanic** | 0.50 | 0.19 | 0.17 | 0.35 | 0.30 | 0.99 | 0.74 | 0.72 | 0.99 | 0.80 | 0.50 | 0.01 |
| **% Less than high school education** | 0.02 | 0.05 | 0.35 | 0.08 | 0.95 | 0.94 | 0.91 | 0.03 | 0.71 | 0.41 | 0.67 | 0.00 |
| **% Renter owner occupied** | 0.00 | 0.00 | 0.16 | 0.07 | 0.09 | 0.06 | 0.11 | 0.59 | 0.92 | 0.98 | 0.72 | 0.00 |
| **Median Household Income** | 0.18 | 0.03 | 0.29 | 0.12 | 0.91 | 0.32 | 0.11 | 0.67 | 0.19 | 0.57 | 0.18 | 0.00 |
| **Median Home Value** | 0.95 | 0.92 | 0.57 | 0.71 | 0.18 | 0.86 | 0.39 | 0.42 | 0.12 | 0.31 | 0.20 | 0.01 |
| **Work** | | | | | | | | | | | | |
| **% Black** | 0.36 | 0.92 | 0.00 | 0.00 | 0.44 | 0.65 | 0.85 | 0.49 | 0.35 | 0.76 | 0.36 | 0.00 |
| **% White** | 0.31 | 0.44 | 0.37 | 0.07 | 0.39 | 0.31 | 0.62 | 0.43 | 0.32 | 0.56 | 0.24 | 0.40 |
| **% Hispanic** | 0.00 | 0.00 | 0.26 | 0.10 | 0.74 | 0.70 | 0.68 | 0.91 | 0.91 | 0.94 | 0.61 | 0.00 |
| **% Less than high school education** | 0.27 | 0.13 | 0.12 | 0.01 | 0.77 | 0.90 | 0.82 | 0.75 | 0.95 | 0.78 | 0.63 | 0.00 |

*[a] p-value was calculated using the Kruskal–Wallis test (non-normally distributed variables, as determined by a Shapiro-Wilk test)*
*[b] p for trend: p-value for trend test using the Jonckheere–Terpstra test.*
*[c] p-value was calculated using the Mann-Whitney U-test (non-normally distributed variables, as determined by a Shapiro-Wilk test)*
*Gray cells signify $p < 0.05$*
*Groups with significant differences using post-hoc Dunn tests:*

- *For complaints per $km^2$: quartiles 2 and 4, and quartiles 3 and 4 for % residents with formal education < high school; quartiles 1 and 4, quartiles 1 and 3, and quartiles 2 and 4 for % renter occupied homes; quartiles 1 and 4, quartiles 2 and 3, quartiles 2 and 4 for % Hispanic workers.*
- *For complaints per population: quartiles 1 and 3, quartiles 1 and 4, quartiles 2 and 4, and quartiles 3 and 4 for % Black workers*
- *For facilities per $km^2$: quartiles 2 and 4, and quartiles 3 and 4 for % residents with formal education < high school*

### 3.3.1 Impact of Gentrification on Complaints

A Mann-Whitney U test revealed that the difference in distribution in complaints between Jan 2014 - May 2023 between gentrifying/non-gentrifying census tracts was not significant ($p = 0.13$). However, differences between the distribution of complaints per $km^2$ (mean values in gentrifying tracts: 7.5, non-gentrifying tracts: 5.1) and complaints per population (1000s) (mean values in gentrifying tracts: 2.0, non-gentrifying tracts: 1.9) in gentrifying versus non-gentrifying census tracts were significant (p= 0.00 and 0.03, respectively).

# 4 Discussion

Our analysis revealed an increase in hotspots of odor complaints, especially after 2018. This observation supports the DDPHE inspectors hypothesis that the rise of social media was enabling more Colorado residents to file odor complaints. However, most of these complaints are occurring in gentrifying neighborhoods which suggests there are likely important differences in complaints by neighborhood which needs to be addressed.

Overall, we observed significant differences in the distribution of all of the environmental justice related variables (except for the percentage of Black residents and the percentage of White workers) considered in census block groups with potentially malodorous facilities compared to those without. Census block groups with such facilities were significantly less privileged (had a higher percentage of non-White residents and workers, had more residents and workers with formal education < high school, had lower median household incomes, had lower house values) than those without. We did not observe similar disparities in the distribution of odor complaints.

However, when we considered the density of odor complaints and facilities in each census block group, we observed significant differences in the distribution of the percentage of residents with formal education < high school across block groups categorized into quartiles based on the number of facilities per $km^2$, although no discernable trend corresponding to the different quartiles was detected. We also observed significant differences in the distribution of the percentage of residents with formal education < high school, percentage of renter occupied homes, and the percentage of Hispanic workers across census block groups categorized into quartiles based on the number of complaints per $km^2$, with clear trends observed for the latter two variables. Specifically, the mean percentage of renter occupied homes was higher for census block groups corresponding to higher intensities of complaints per $km^2$; while no clear pattern was observed in the mean percentage of Hispanic workers across quartiles: the percentage increased between quartiles 1 and 2, and then decreased over quartiles 2, 3 and 4.

Our findings reveal clear disparities in the locations of facilities with less privileged census block groups disproportionately burdened with such facilities. The distribution in complaints was less clear. There could be several reasons the lack of clear patterns in the distribution of complaints: 1) It may be that more privileged individuals feel more empowered to register complaints, 2) It is possible that some communities are more aware of complaint procedures than others, 3) It is likely that only certain facilities release problematic odors and the complaints are in response to this small number of facilities, 4) meteorology could be a factor, with odors spreading downwind from the source depending on the wind direction and speed (we saw this in our paper - that I referenced above). From the topic modeling exercise, for example, we note that a large number of complaints recorded were in response to foul odors from the Purina factory that produces dog and cat food. 4) Meteorological factors can play a key role in dispersing odors over Denver, 5)

Or a combination of the above factors. More research is needed to understand our findings on the distribution of complaints.

Importantly, our study suggests disparities in not just traditional residence based environmental justice related variables, but in workplace based environmental justice variables, as well. Our work thus points to the need to broaden our understanding of the structural racism forces that shape disparities from factors such as residential segregation to that of access to transportation and other less commonly studied dimensions of structural racism which also lead to disproportionate exposures at work by less-resourced populations. Finally, we observed that the density of complaints were higher in gentrifying neighborhoods compared to others. To the best of our knowledge, this study presents the first comprehensive analysis of the distribution of odor facilities and complaints across Denver through an environmental justice lens. It also analyzes disparities in each of these metrics of odor using both residence and workplace based environmental justice variables.

A key limitation of this study is the use of the locations of odor facilities and complaints as a proxy for odor. However, as the wide range of chemical compounds that emit odor are not routinely measured, and many cannot be measured as concentrations are below the detection limit of existing instruments; the proxies for odor that we consider in this study have been used previously in the literature[19,20]. The odor management facilities considered in this study were those required to submit odor management plans as of 2023. We do not have data of such facilities in previous years. We considered several environmental justice variables but did not exhaustively consider the many relative variables of interest. Multiple comparisons may increase the occurrence of significant findings by chance. Future work could include additional variables and consider multiple comparison issues when using a larger number of metrics.

Nevertheless, our study adds to the growing literature on disparities associated with odor and odor complaints. These results have implications for future studies on environmental justice and odor exposure and suggest the need for further research using detailed exposure assessment in other locations.

# Conflict of Interest

None

# Acknowledgements

The authors are grateful to staff at DDPHE for comments on this document

# Supplementary Information

## Evaluating the Environmental Justice Dimensions of Odor in Denver, Colorado

Priyanka N. deSouza[1,2*], Amanda Rees[1], Emilia Oscilowicz[1], Brendan Lawlor[3], William Obermann[3], Katherine Dickinson[4], Lisa M. McKenzie[4], Sheryl Magzamen[5,6], Shelly Miller[7], Michelle L. Bell[8]

1: Department of Urban and Regional Planning, University of Colorado Denver, Denver CO, 80202, USA
2: CU Population Center, University of Colorado Boulder, Boulder, CO 80302, USA
3: Denver Department of Public Health and Environment, Denver CO, 80202, USA
4: Department of Environmental and Occupational Health, Colorado School of Public Health, University of Colorado Anschutz, Aurora, CO, 80045, USA
5: Department of Environmental and Radiological Health Sciences, Colorado State University, Fort Collins, CO, USA
6: Department of Epidemiology, Colorado School of Public Health, Colorado State University, Fort Collins, CO, USA
6: Dept. of Mechanical Engineering and Program of Environmental Engineering, Univ. of Colorado Boulder, Boulder, CO 80309
7: School of the Environment, Yale University, New Haven, CT, 06520-8354, USA

*: priyanka.desouza@ucdenver.edu

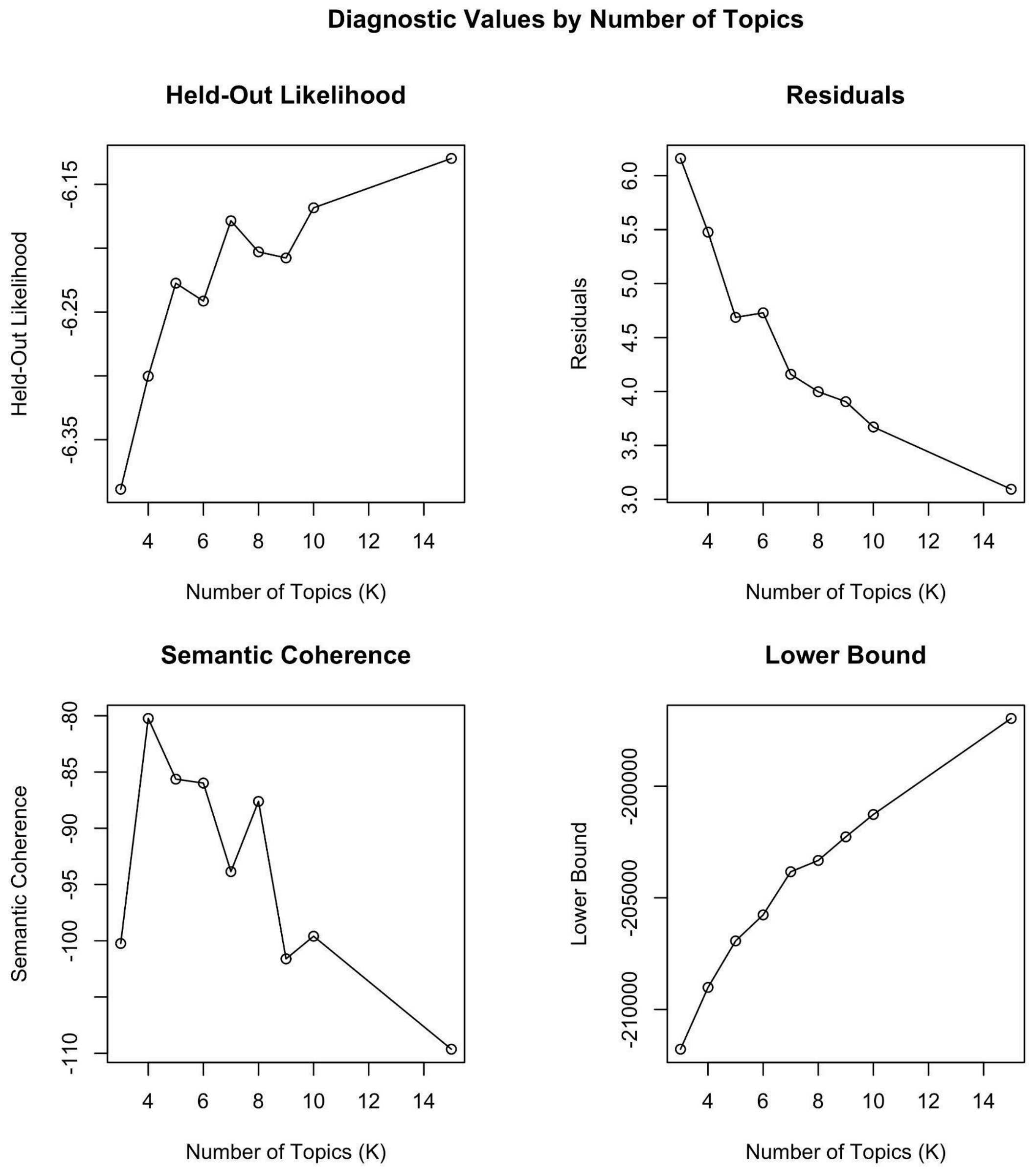

***Figure S1**: Diagnostics to determine the number of topics.*

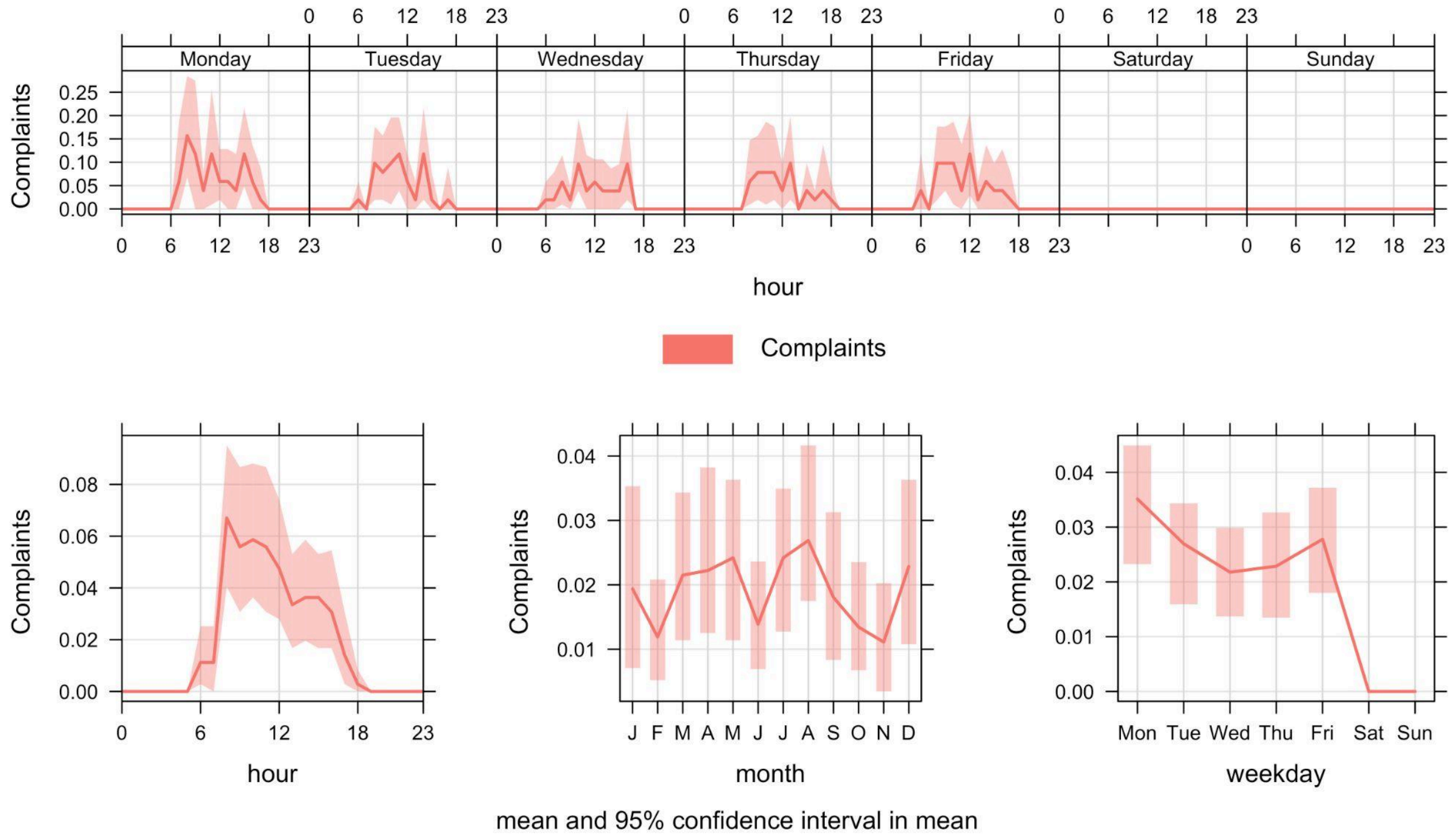

***Figure S2****: Time variation of odor complaints in 2014 over an average day, month and day-of-the-week*

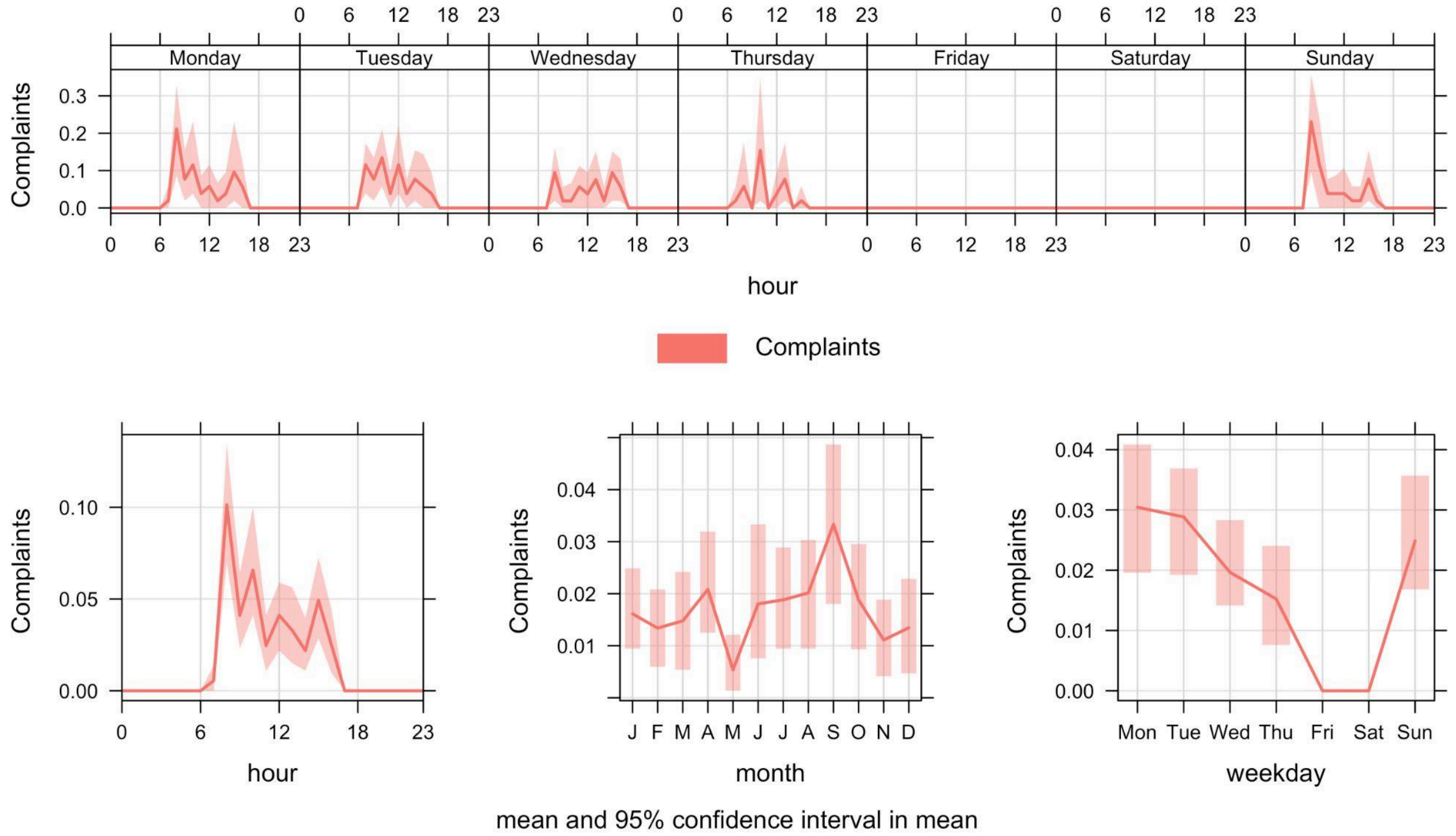

***Figure S3****: Time variation of odor complaints in 2015 over an average day, month and day-of-the-week*

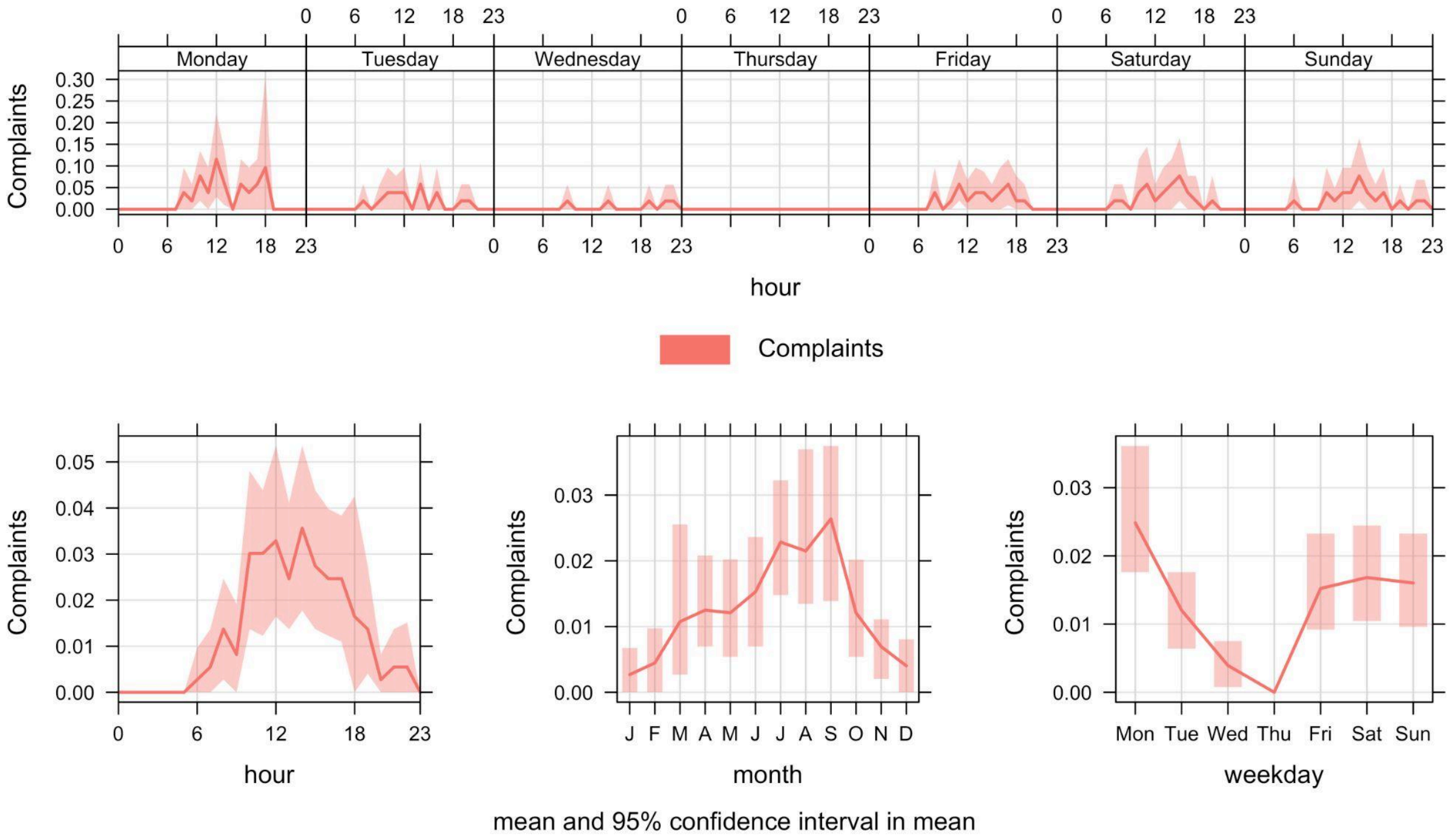

***Figure S4**: Time variation of odor complaints in 2016 over an average day, month and day-of-the-week*

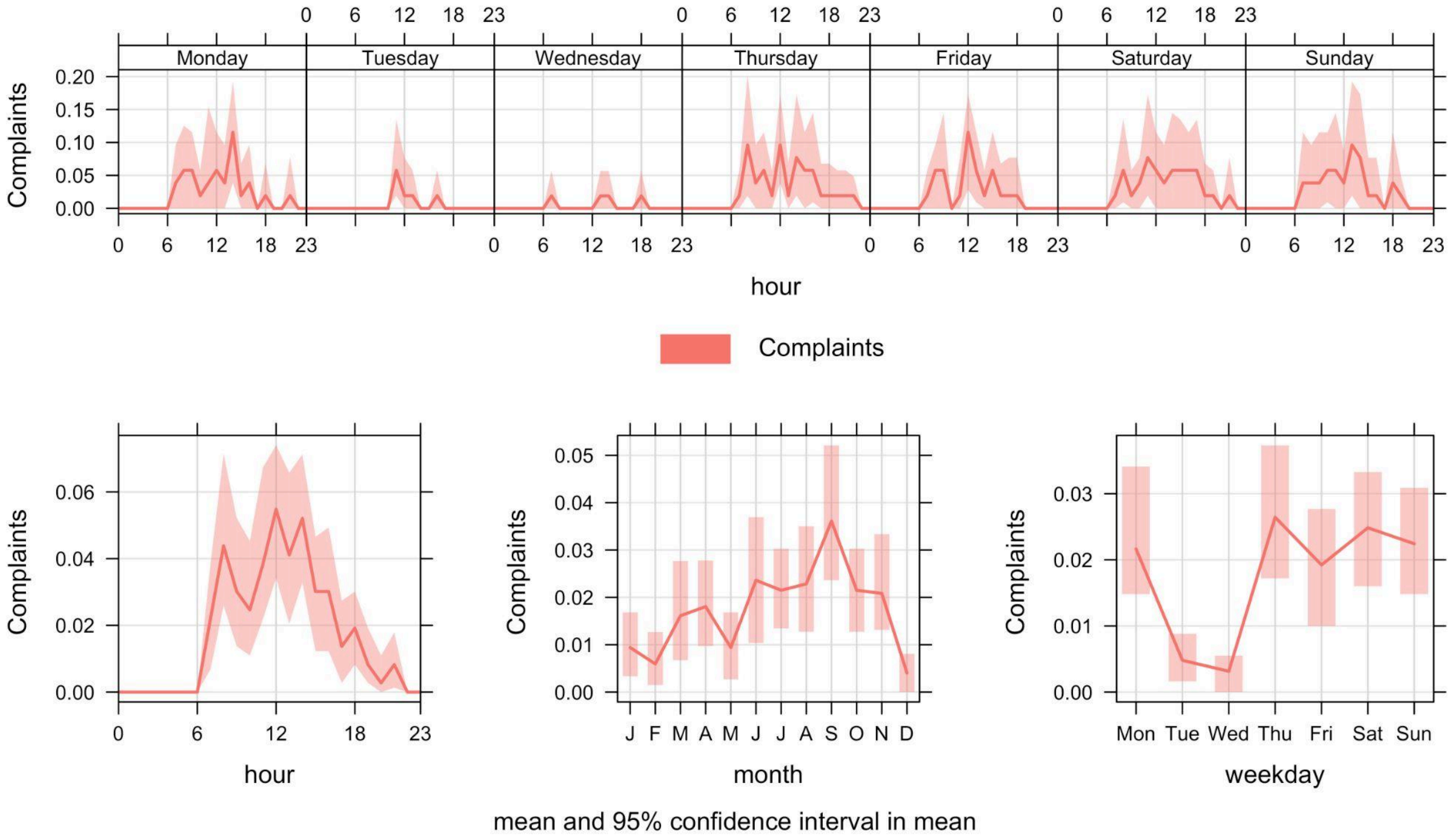

***Figure S5**: Time variation of odor complaints in 2017 over an average day, month and day-of-the-week*

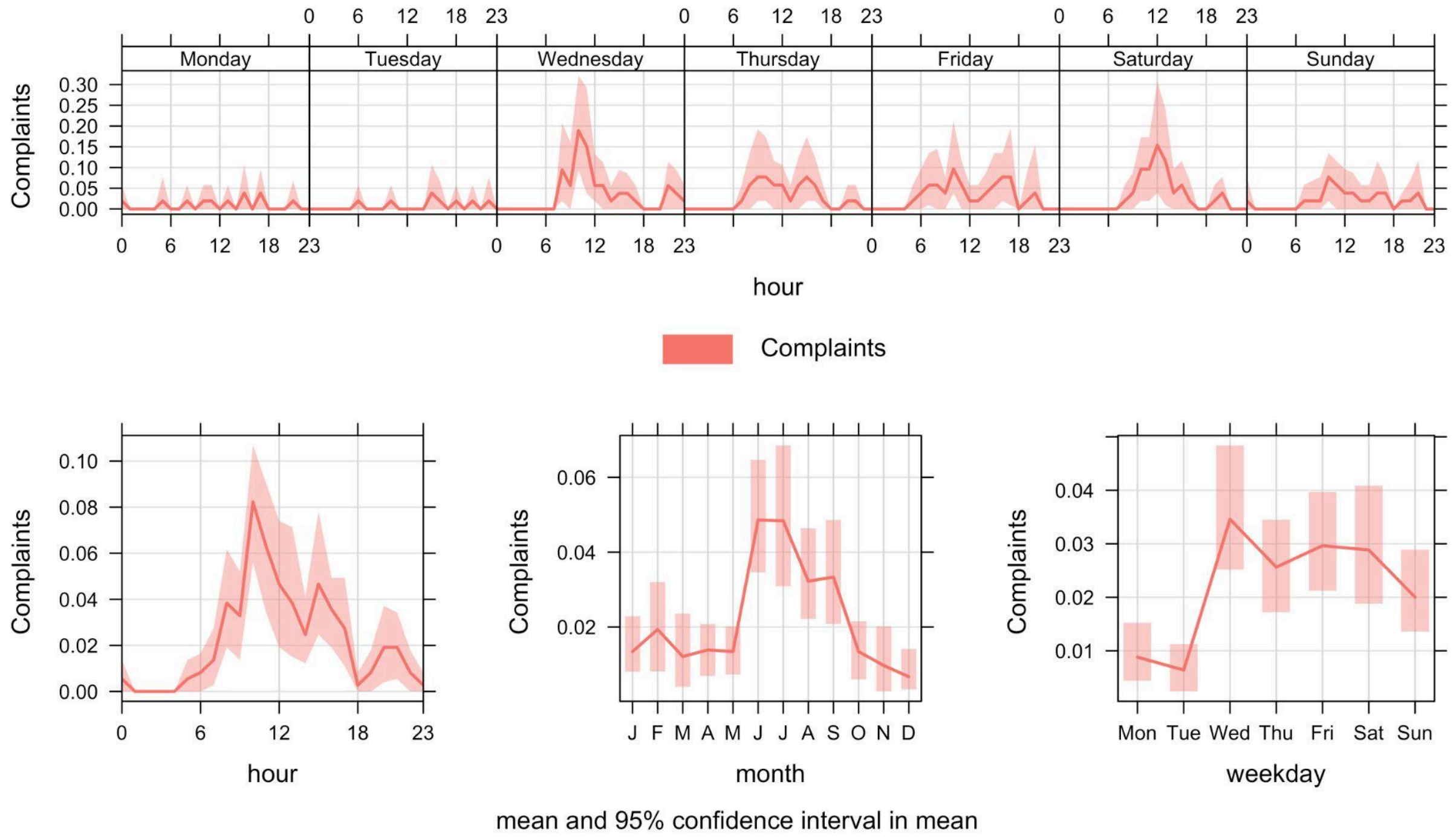

***Figure S6***: *Time variation of odor complaints in 2018 over an average day, month and day-of-the-week*

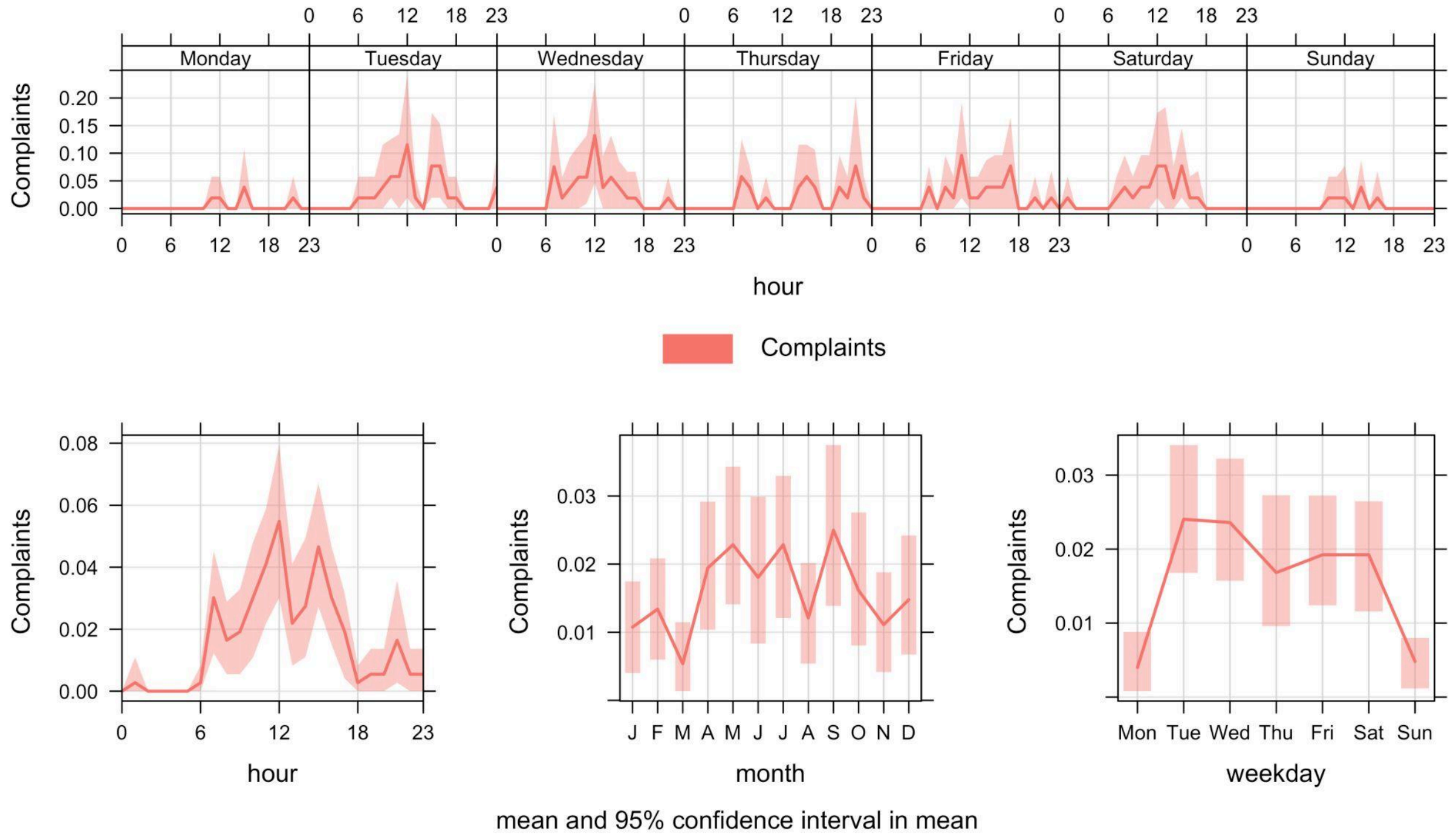

***Figure S7***: *Time variation of odor complaints in 2019 over an average day, month and day-of-the-week*

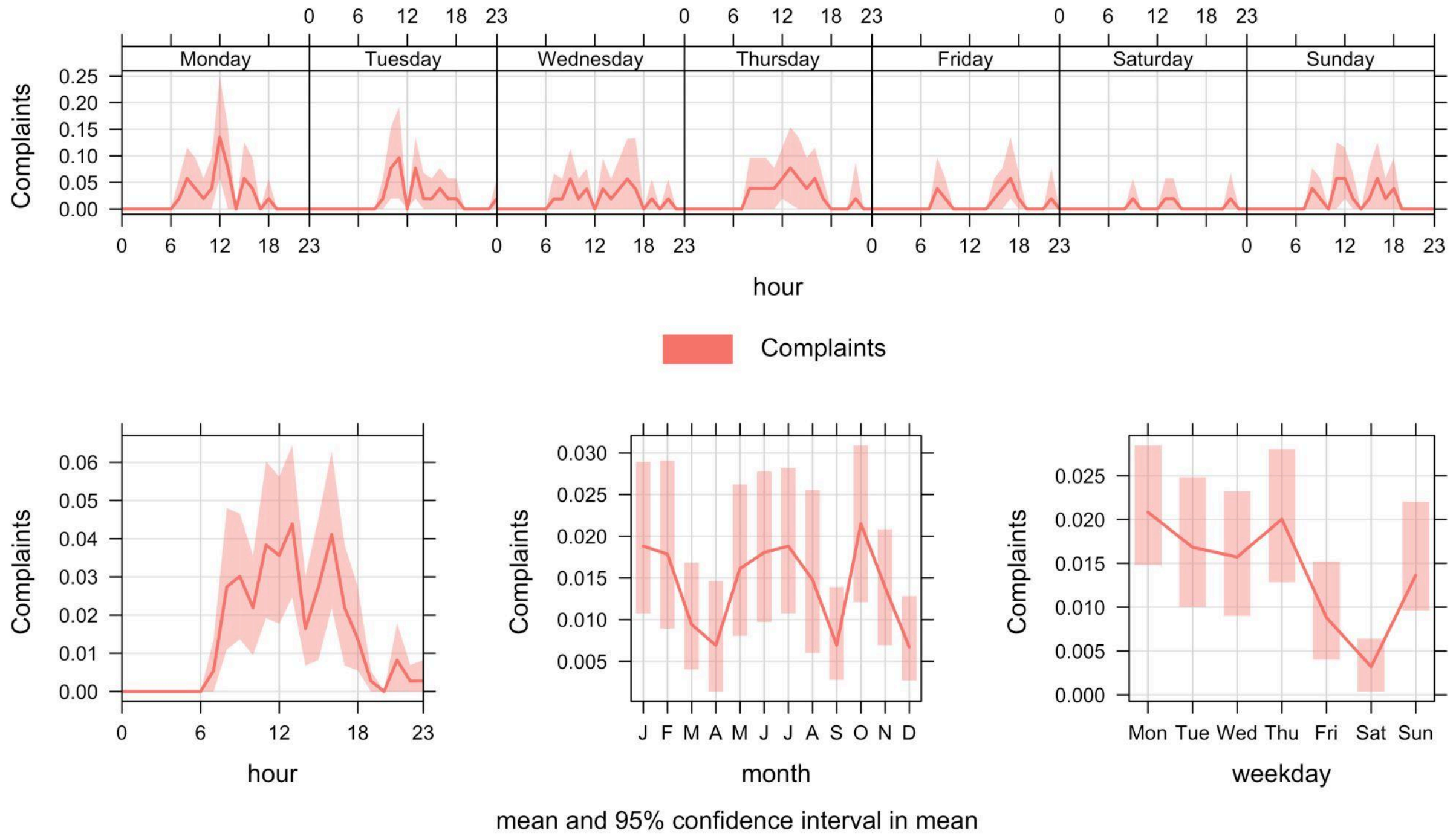

***Figure S8***: *Time variation of odor complaints in 2020 over an average day, month and day-of-the-week*

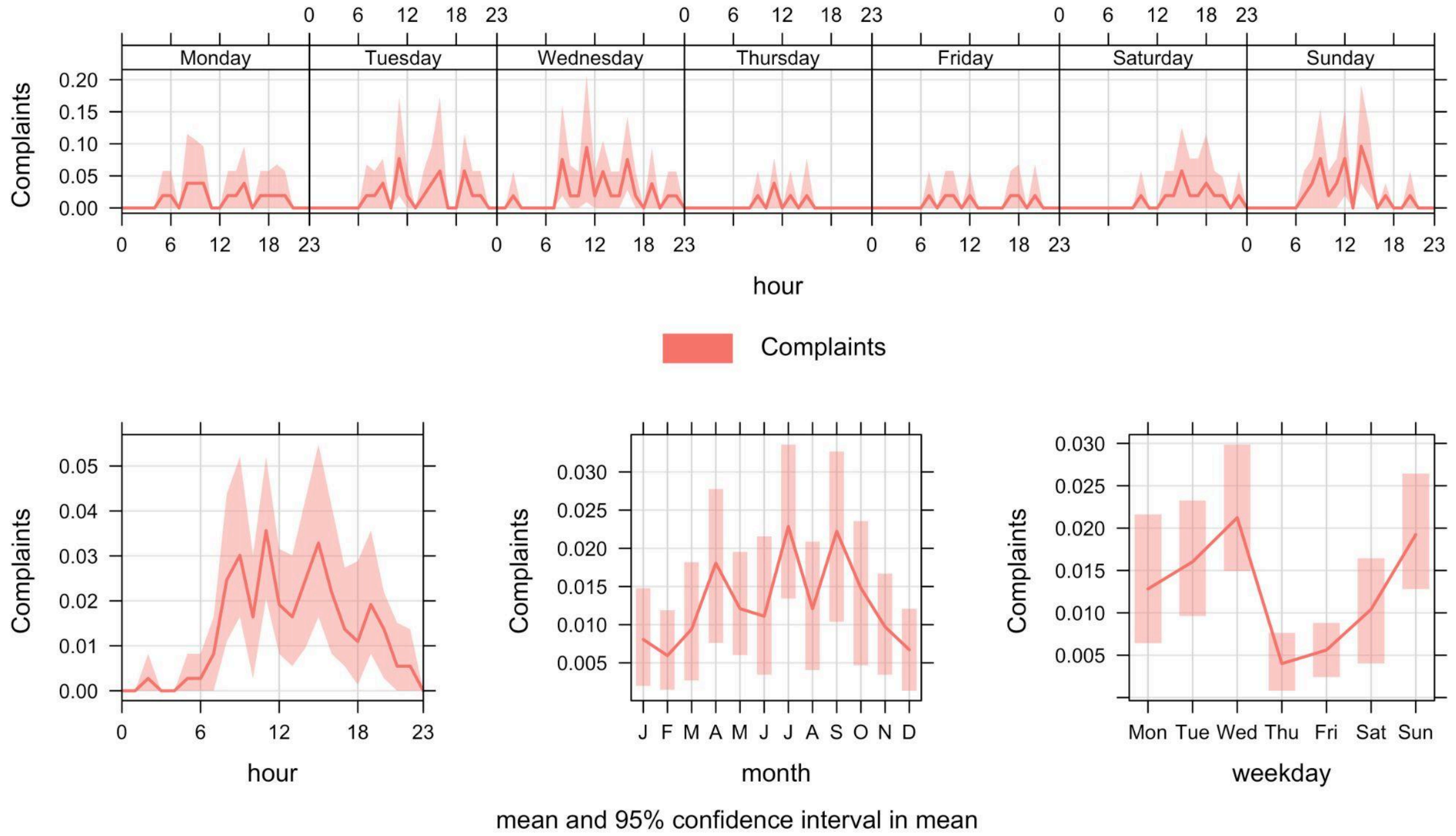

***Figure S9***: *Time variation of odor complaints in 2021 over an average day, month and day-of-the-week*

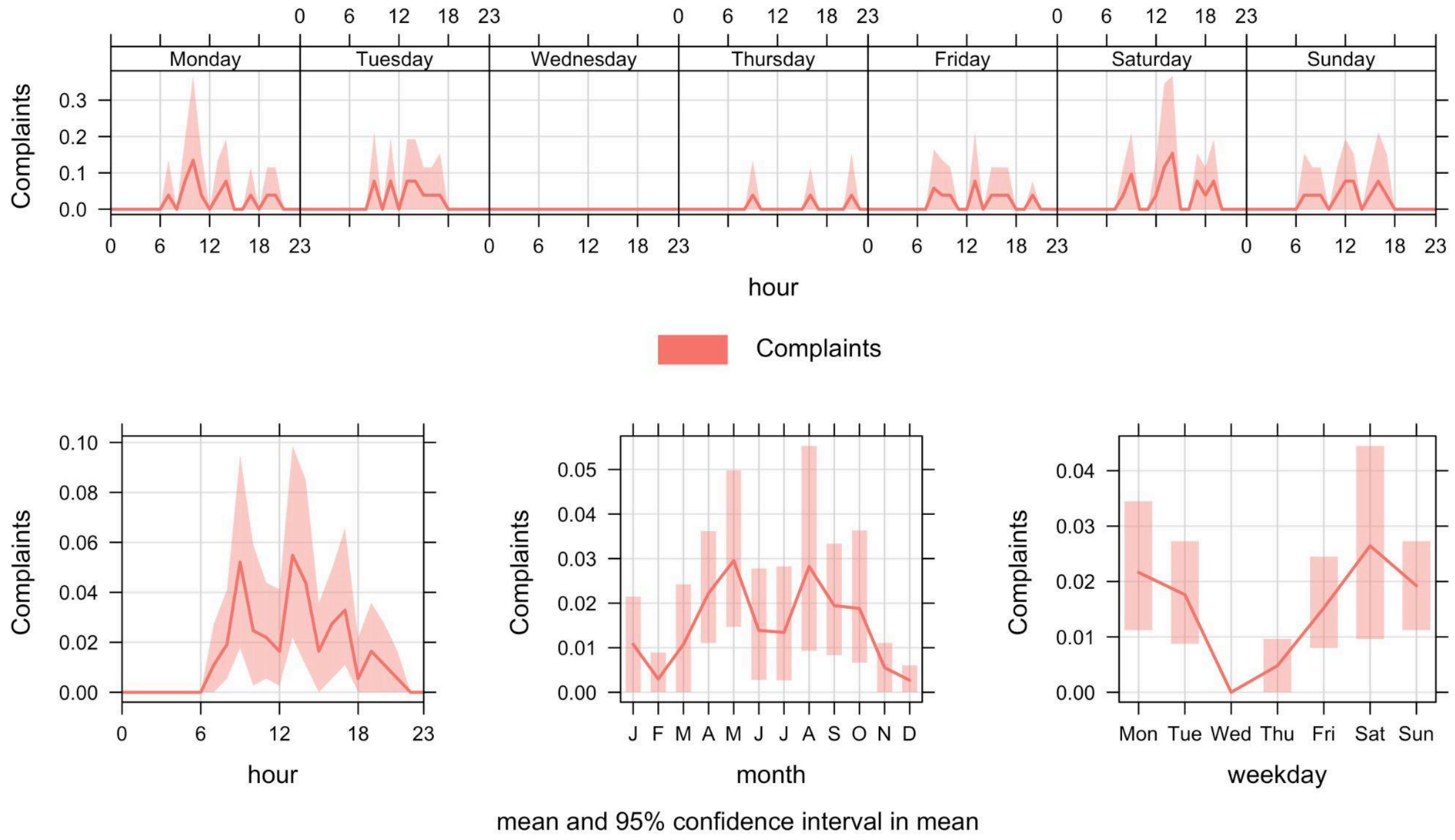

***Figure S10**: Time variation of odor complaints in 2022 over an average day, month and day-of-the-week*

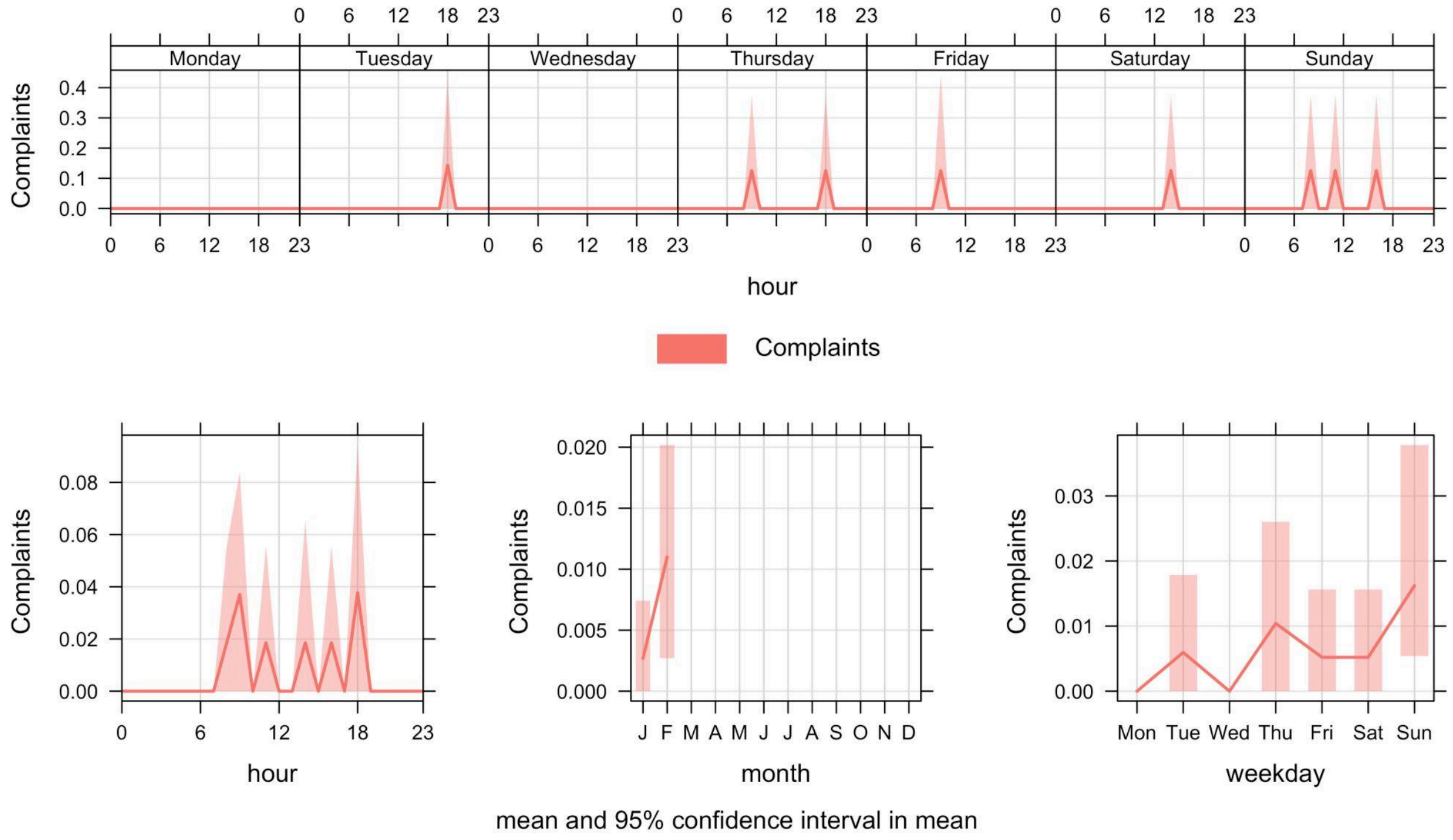

***Figure S11**: Time variation of odor complaints in 2023 over an average day, month and day-of-the-week*

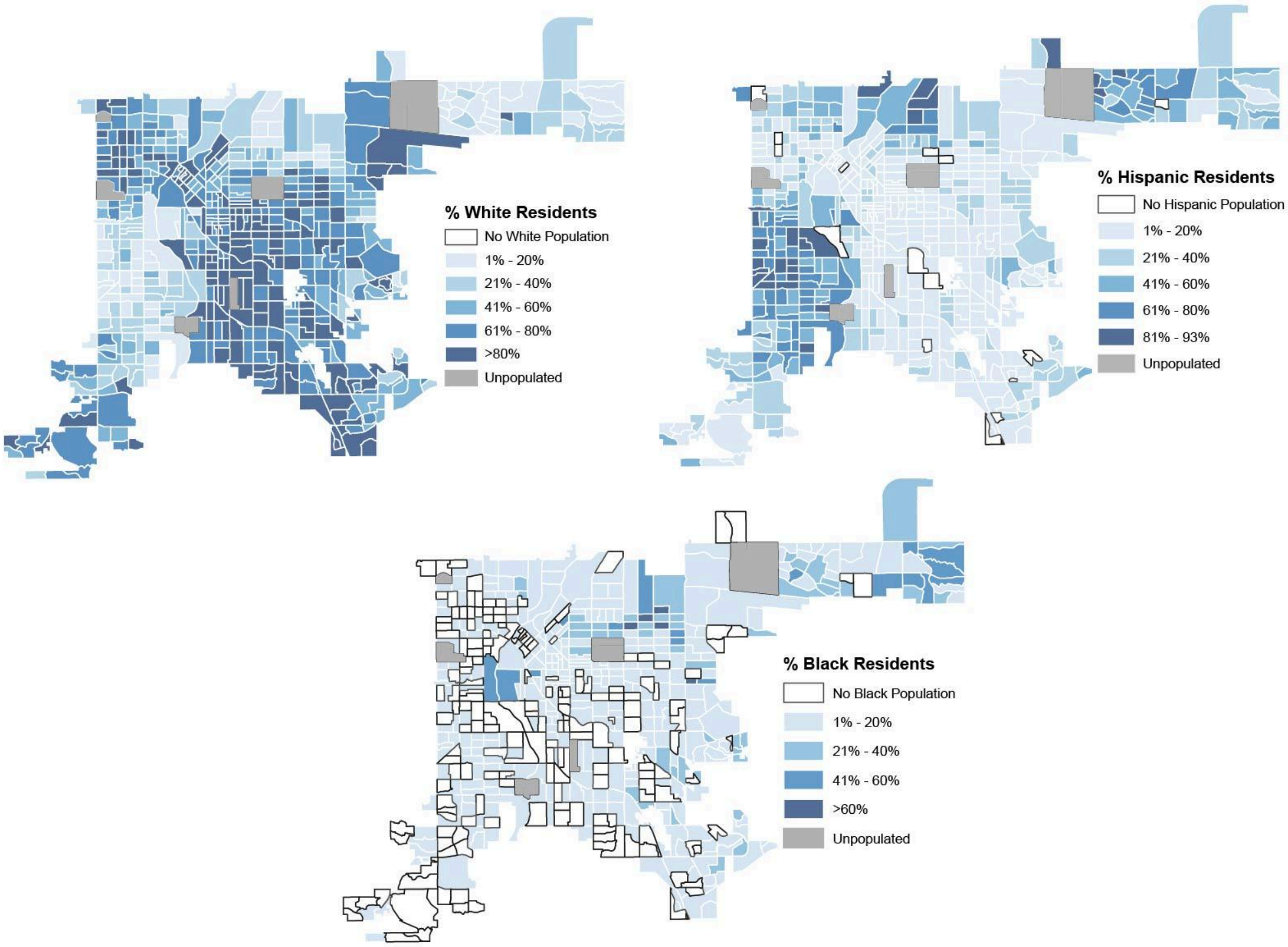

***Figure S12***: *Race related environmental justice variables based on place of residence: A) % of white residents, B) % of Black residents, C) % of Hispanic residents at the blockgroup level for Denver*

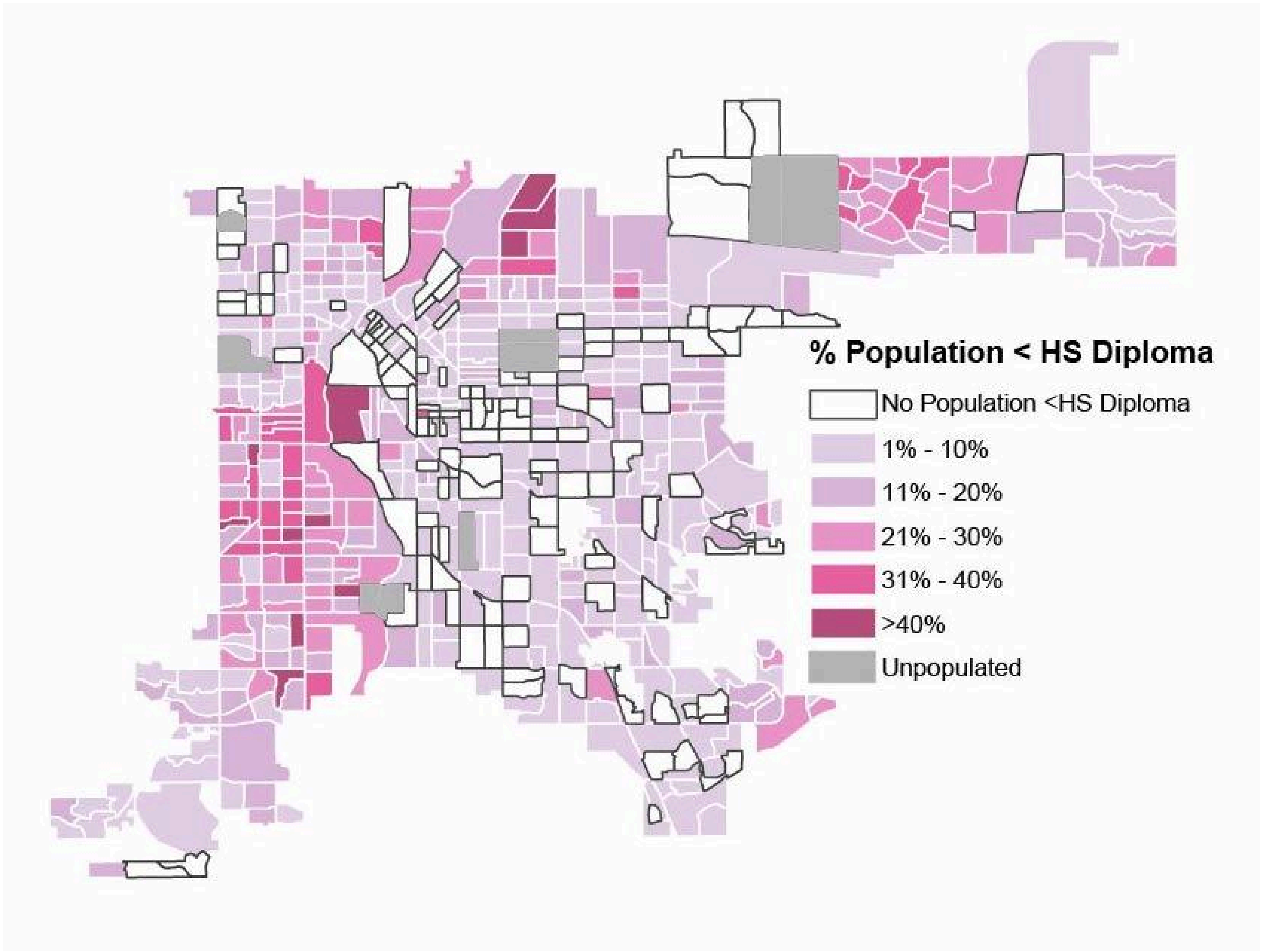

***Figure S13***: *Education related environmental justice variables based on place of residence: % of residents with education < high school at the blockgroup level in Denver.*

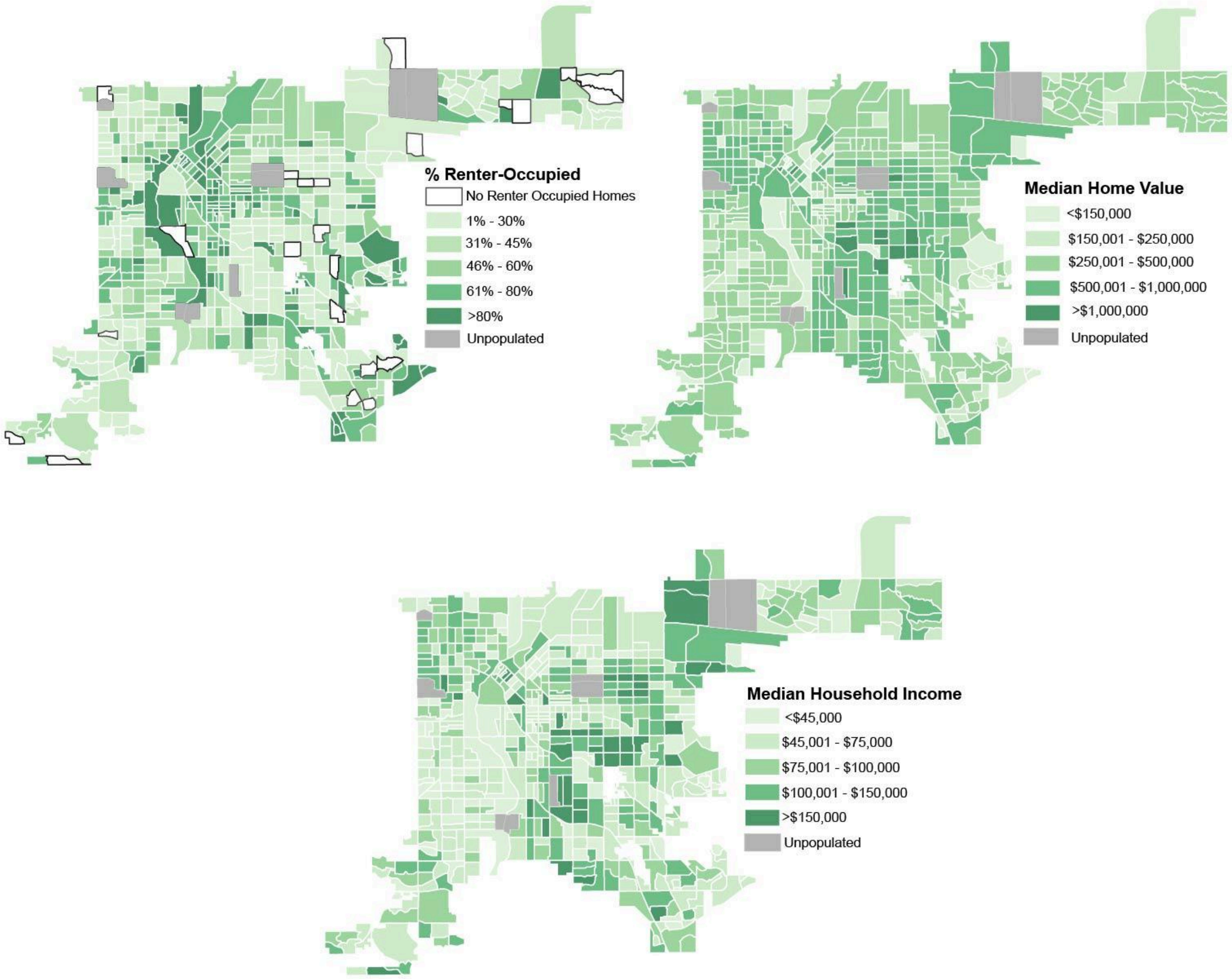

***Figure S14****: Wealth/income related environmental justice variables based on place of residence: % of residents with education < high school at the blockgroup level in Denver.*

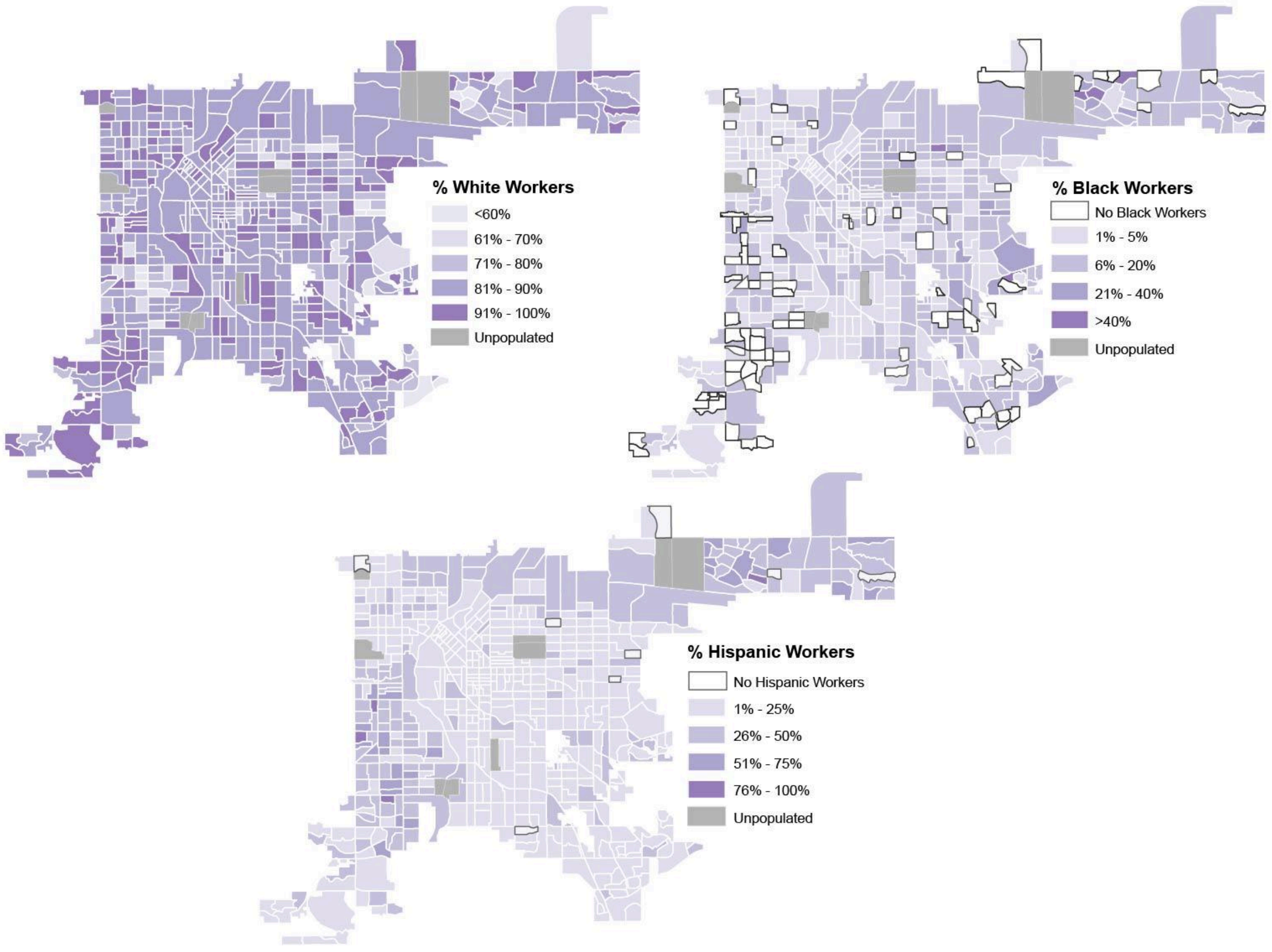

***Figure S15**: Race related environmental justice variables based on place of work: A) % of white workers, B) % of Black workers, C) % of Hispanic workers at the blockgroup level for Denver*

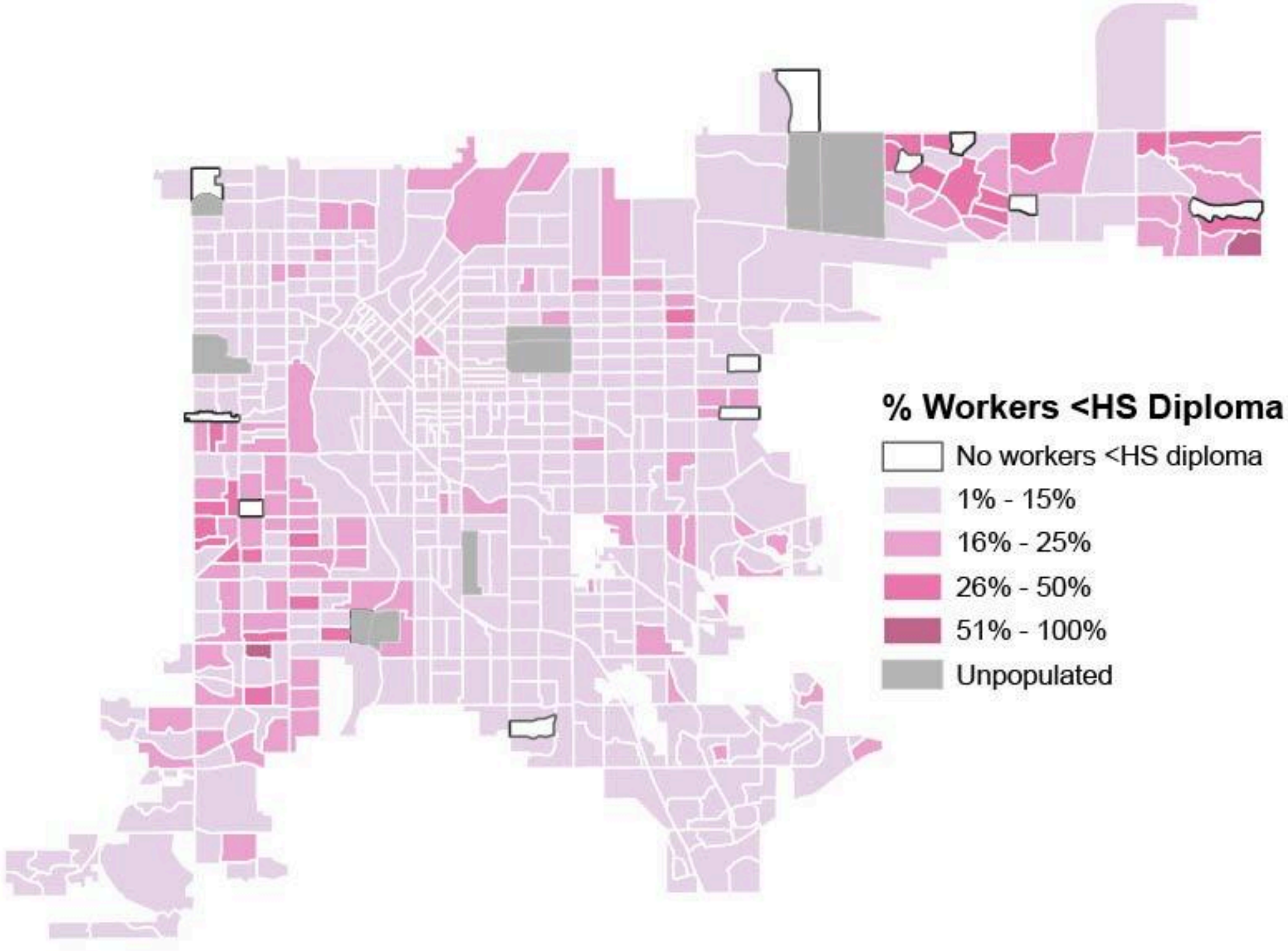

***Figure S16:*** *Education related environmental justice variables based on place of work: % of workers with education < high school at the blockgroup level in Denver.*

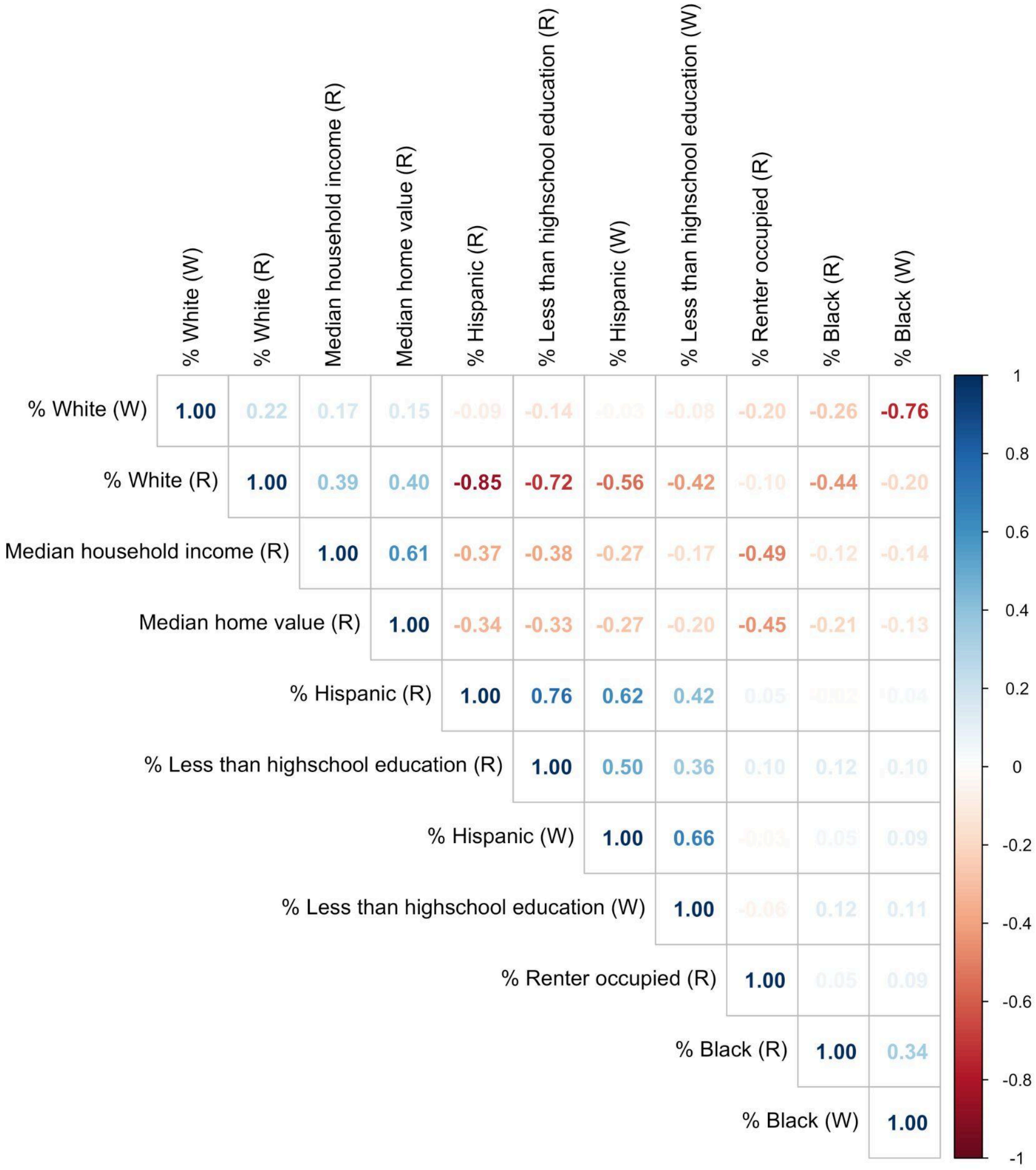

***Figure S17**: Pairwise Pearson correlations between the residential (R) and work-based (W) environmental justice related variables used in this study across Denver. Note that correlations were calculated for the 560 (of 571) census blockgroups which had a non-zero number of residents and workers.*

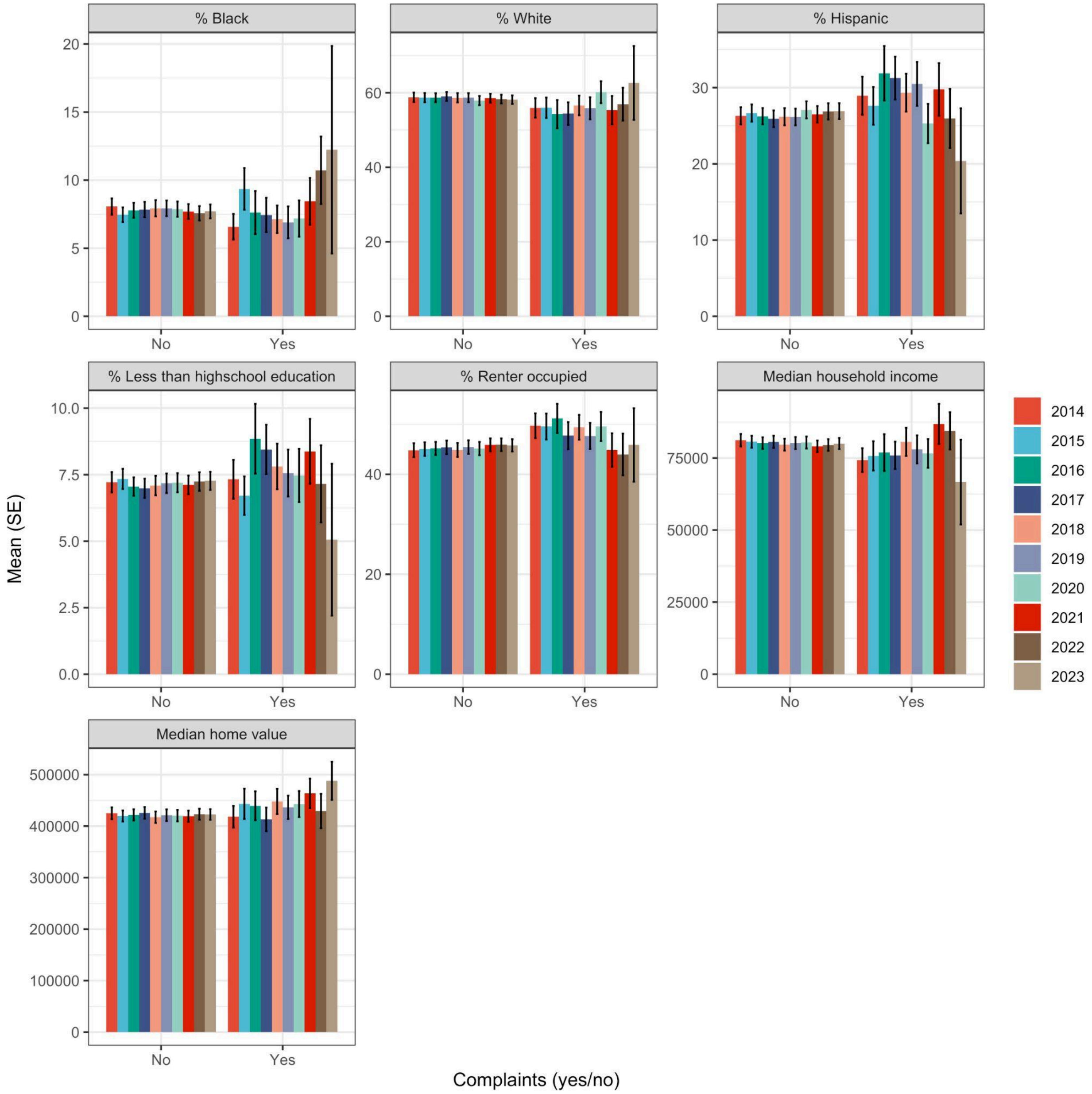

***Figure S18****: Distribution (Mean  with SE error bars) of residential based environmental justice related variables for census blockgroups with and without complaints registered for each year between 2014-2022 in Denver.*

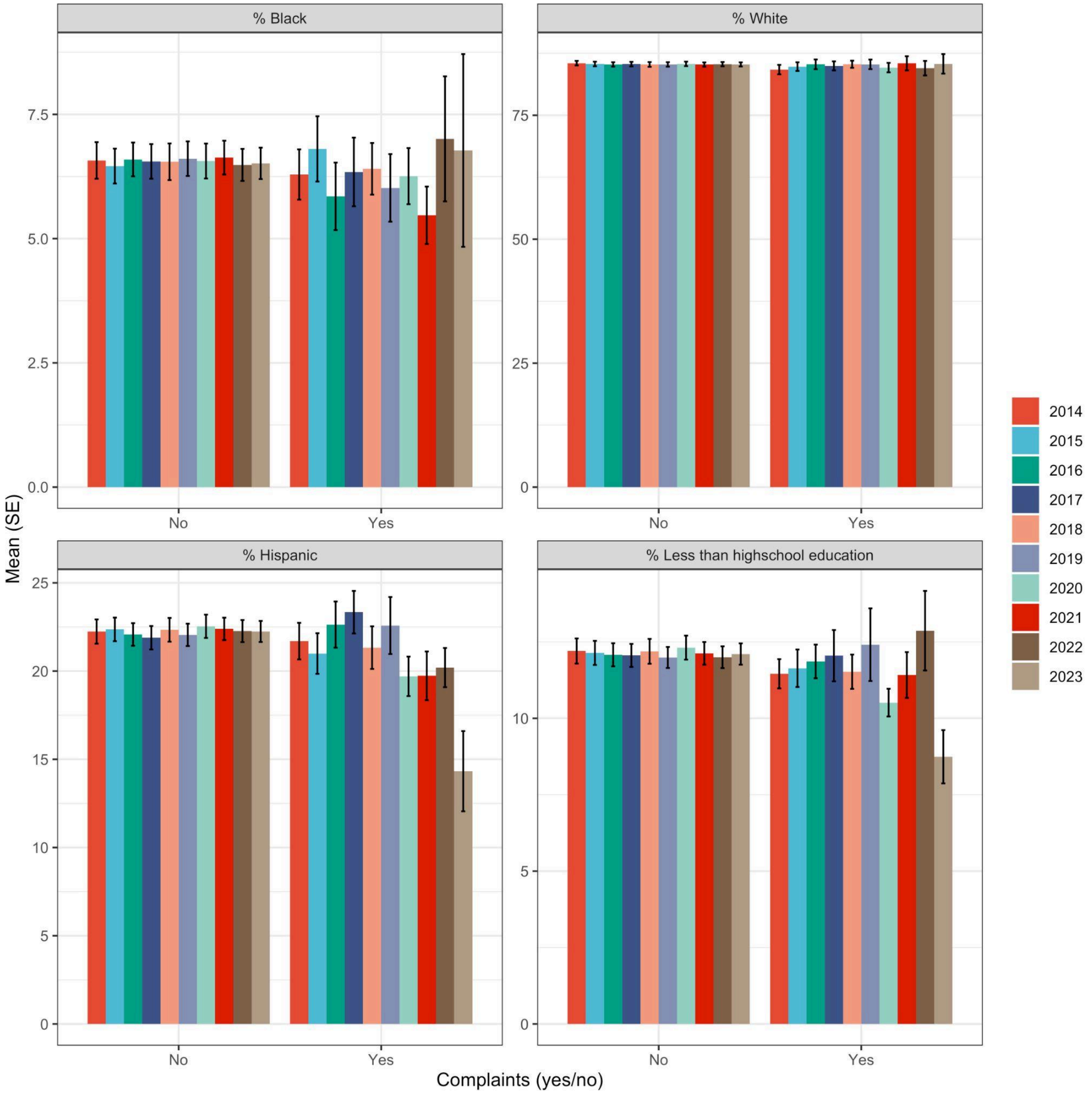

***Figure S19***: *Distribution (Mean with SE error bars) of workplace-based environmental justice related variables for census blockgroups with and without complaints registered for each year between 2014-2022 in Denver.*

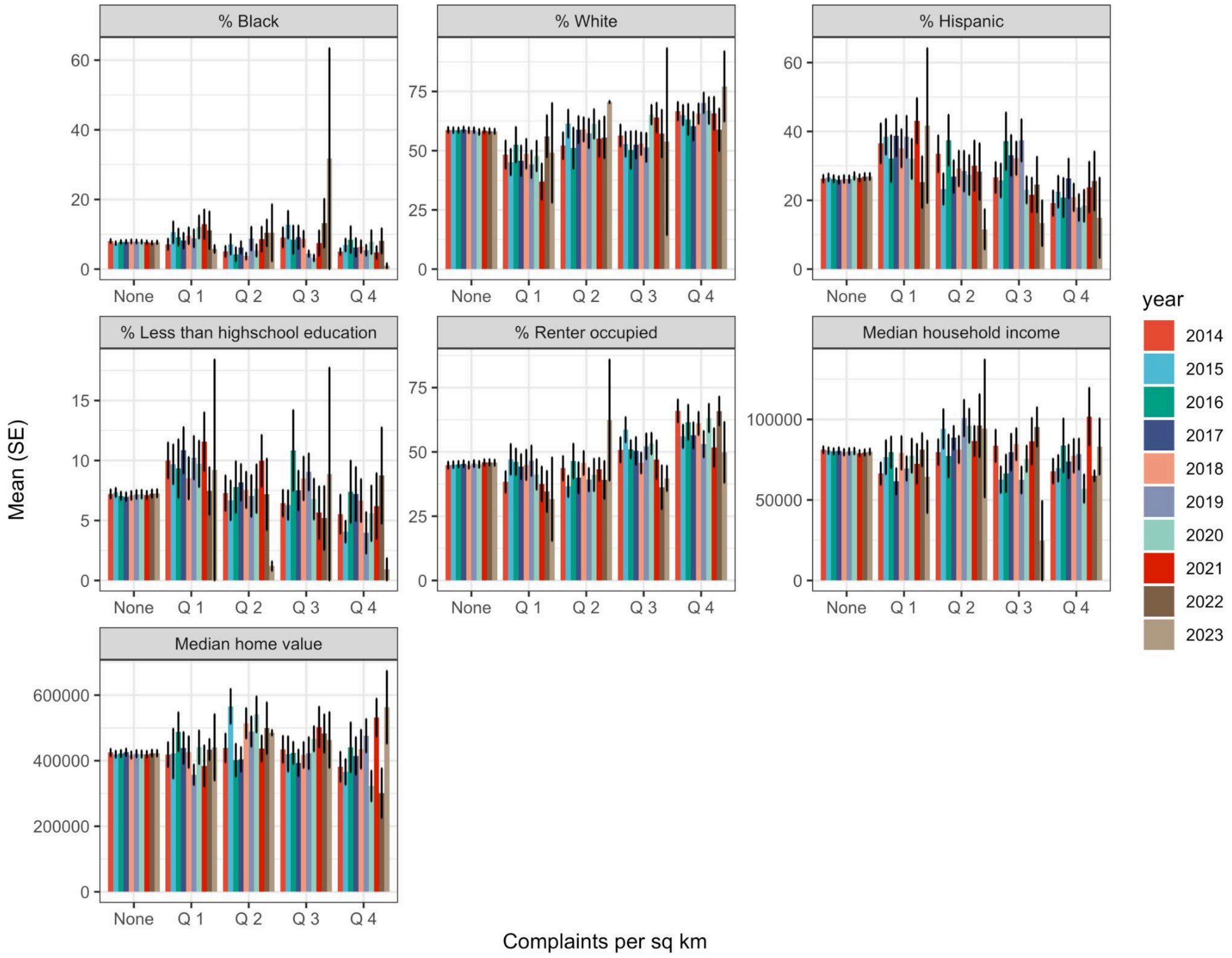

***Figure S20****: Distribution (Mean with SE error bars) of residential based environmental justice related variables over census blockgroups belonging to different quartiles based on the number of complaints registered every year/km$^2$ of blockgroup in Denver.*

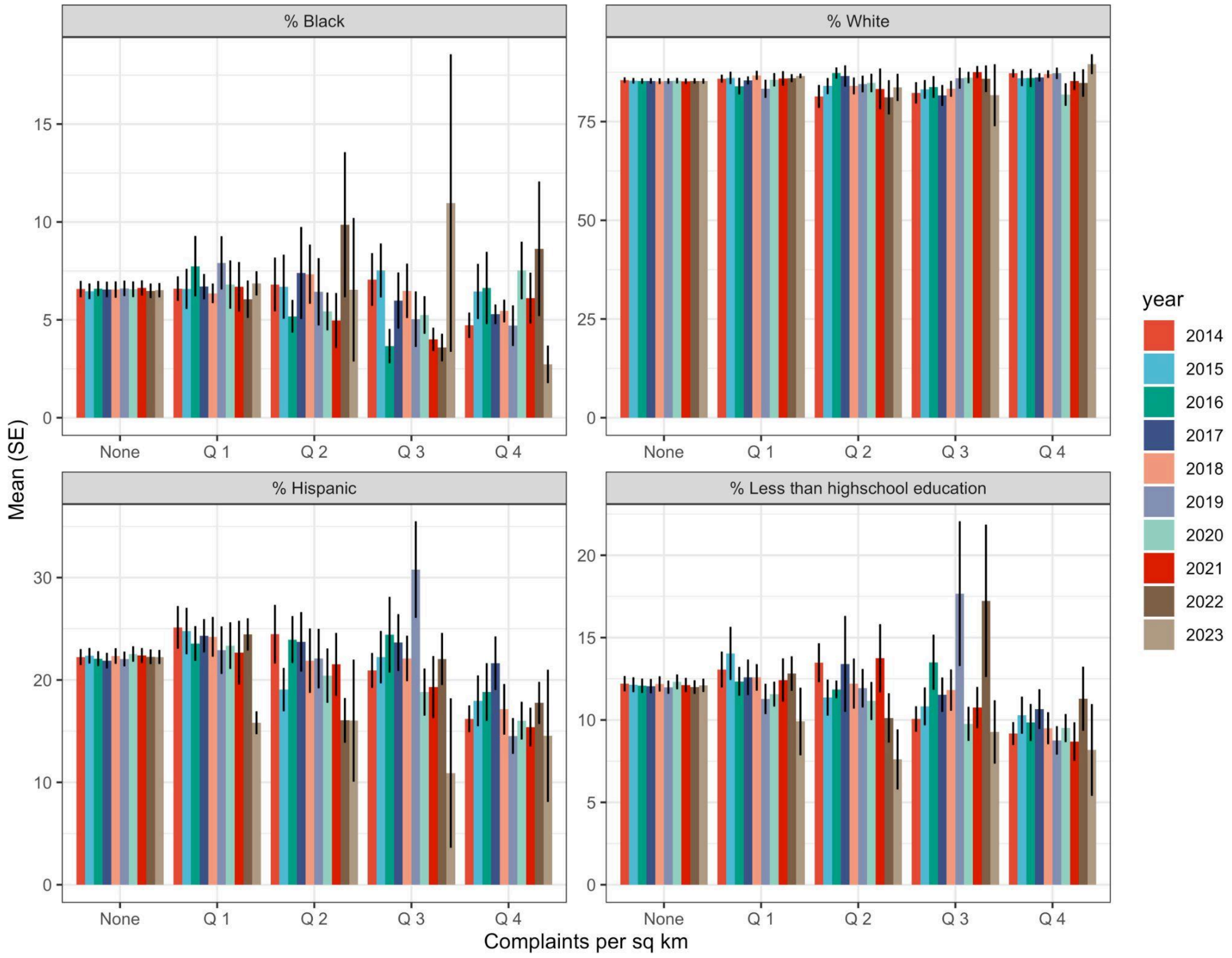

***Figure S21****: Distribution (Mean with SE error bars) of work based environmental justice related variables over census blockgroups belonging to different quartiles based on the number of complaints registered every year/km$^2$ of blockgroup in Denver.*

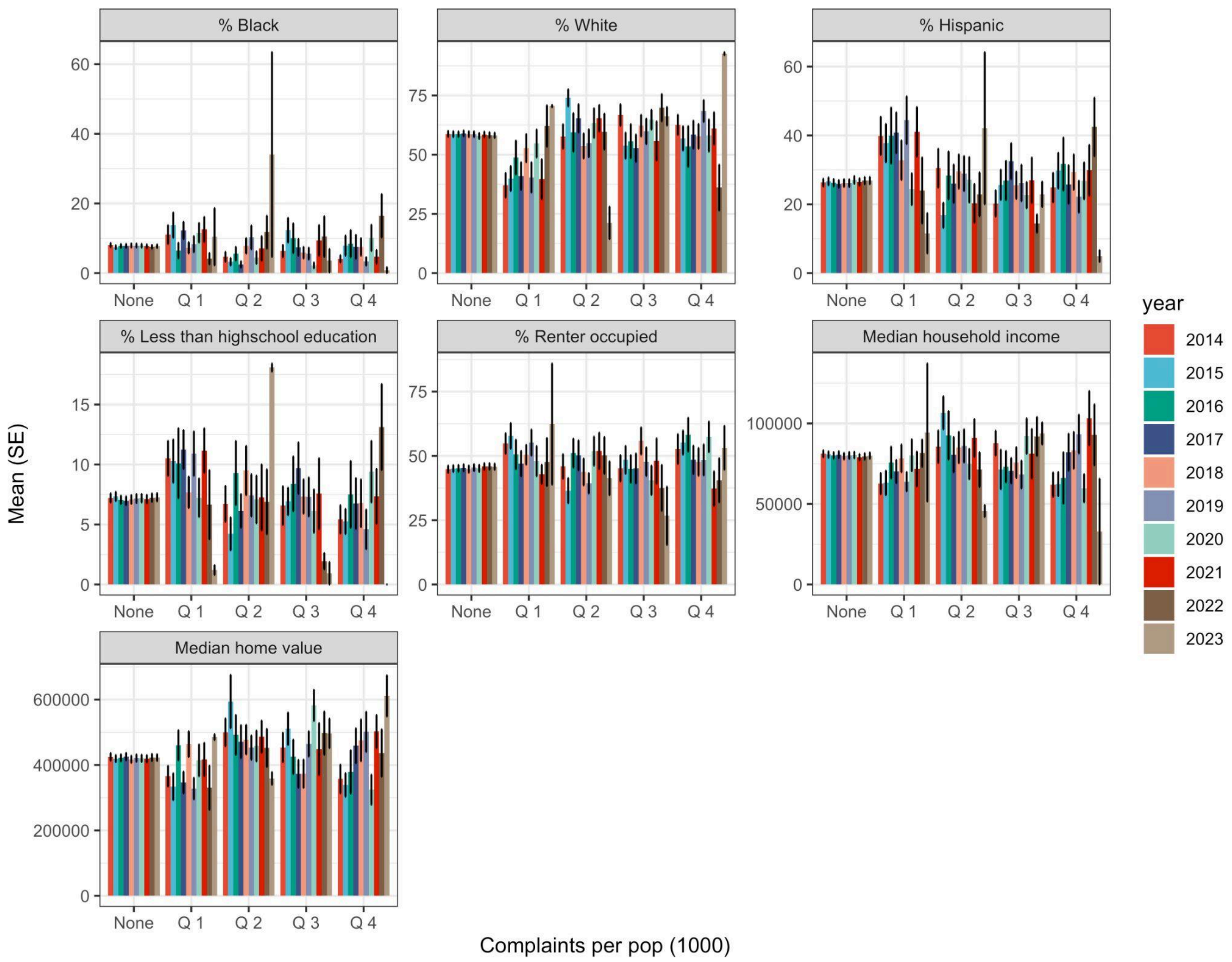

***Figure S22**: Distribution (Mean with SE error bars) of residential based environmental justice related variables over census blockgroups belonging to different quartiles based on the number of complaints registered every year/population of blockgroup in Denver.*

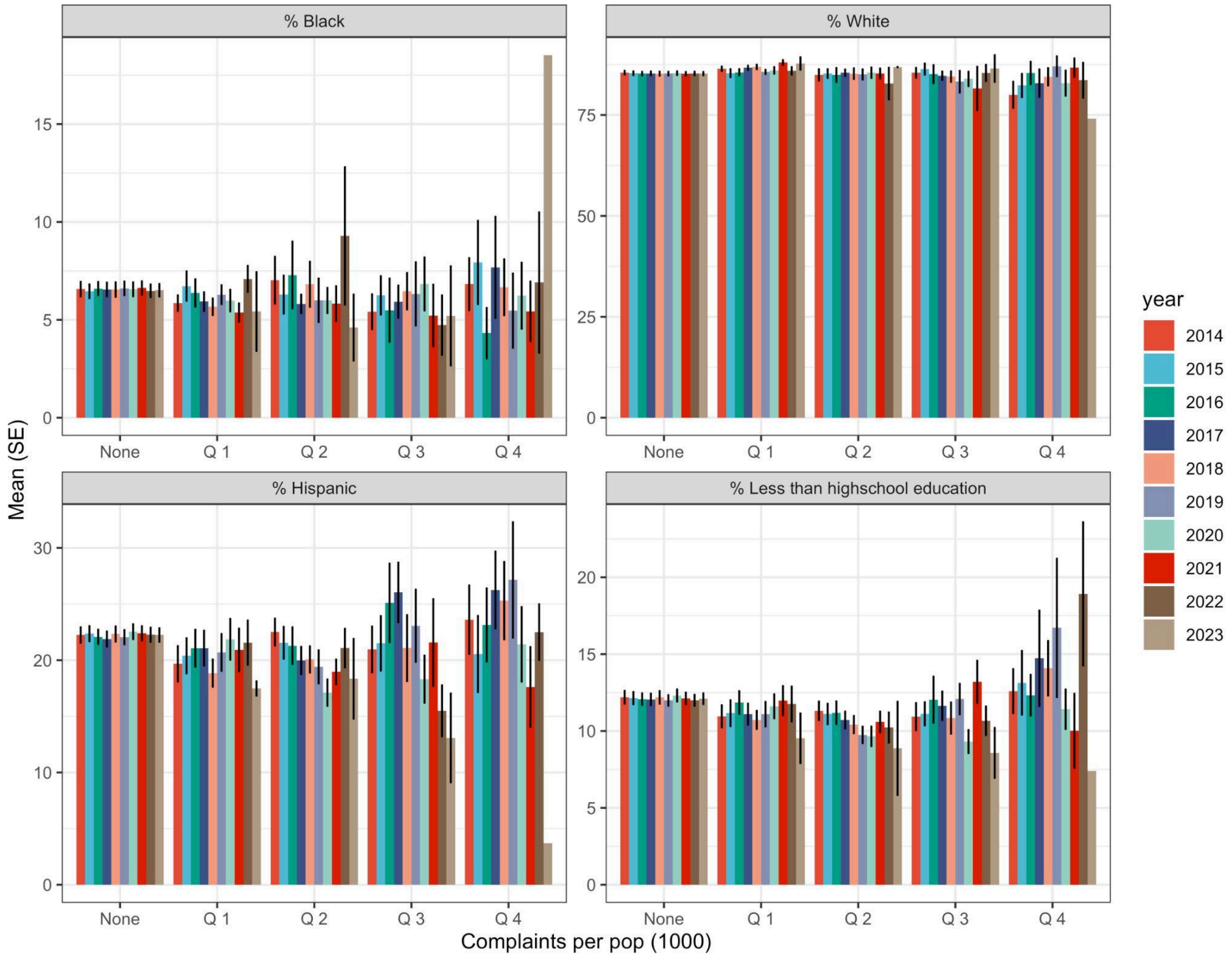

***Figure S23***: *Distribution (Mean with SE error bars) of work based environmental justice related variables over census blockgroups belonging to different quartiles based on the number of complaints registered every year/population of blockgroup in Denver.*

# Tables

***Table S1**: Descriptive statistics of complaints registered every year between Jan 2014 - May 2023*

| Year | Number of Complaints | Number of Unique Census Block Groups |
|---|---|---|
| 2014 | 172 | 112 |
| 2015 | 149 | 95 |
| 2016 | 114 | 58 |
| 2017 | 153 | 97 |
| 2018 | 201 | 114 |
| 2019 | 153 | 88 |
| 2020 | 124 | 80 |
| 2021 | 114 | 55 |
| 2022 | 134 | 37 |
| 2023 | 8 | 8 |

***Table S2**: Range of each exposure metric corresponding to each quartile.*

| **Exposure Metrics** | | **Quartile 1** | **Quartile 2** | **Quartile 3** | **Quartile 4** |
|---|---|---|---|---|---|
| **Residential** | | | | | |
| **Complaints per km$^2$** | N | 87 | 87 | 87 | 87 |
| | Range | 0.18-2.82 | 2.86-5.51 | 5.53-11.18 | 11.22-185.69 |
| **Complaints per population (1000s)** | N | 87 | 87 | 87 | 87 |
| | Range | 0.27-0.97 | 0.98-1.74 | 1.74-3.27 | 3.28-140.24 |
| **Facilities per km$^2$** | N | 11 | 11 | 11 | 11 |
| | Range | 0.71-2.74 | 2.79-5.55 | 5.65-8.19 | 8.29-19.37 |
| **Facilities per population (1000s)** | N | 11 | 11 | 11 | 11 |
| | Range | 0.36-1.41 | 1.56-2.79 | 3.09-6.16 | 6.39-384.62 |
| **Mean time to resolution** | N | 87 | 87 | 86 | 87 |
| | Range | 0-1.5 | 1.6-5.0 | 5.3-13.7 | 14.0-257.0 |
| **Work** | | | | | |
| **Complaints per km$^2$** | N | 88 | 87 | 87 | 88 |
| | Range | 0.18-2.82 | 2.86-5.24 | 5.51-11.02 | 11.18-185.69 |
| **Complaints per population (1000s)** | N | 88 | 88 | 86 | 88 |
| | Range | 0.04-2.49 | 2.51-9.35 | 9.43-27.40 | 28.99-1000.0 |
| **Facilities per km$^2$** | N | 12 | 11 | 11 | 11 |
| | Range | 0.71-2.79 | 3.03-5.65 | 5.84-8.29 | 8.63-19.37 |
| **Facilities per population (1000s)** | N | 12 | 11 | 11 | 11 |
| | Range | 0.19-1.77 | 1.90-3.23 | 3.30-6.80 | 7.91-58.82 |
| **Mean time to resolution** | N | 88 | 88 | 87 | 86 |
| | Range | 0-1.6 | 1.7-5.3 | 5.4-14.0 | 14.3-257.0 |

*N: number of unique census blockgroups*

**Table S3**: *Results from statistical tests to compare the distribution of environmental justice related variables by exposure intensity based on different exposure metrics based on the number of facilities at the census blockgroup level in Denver*

| | Marijuana Facilities | | | | | Non-marijuana Facilities |
|---|---|---|---|---|---|---|
| | Per sq km | | Per Population | | Yes/ No | Yes/ No |
| | p-value[a] | p-value trend[b] | p-value[a] | p-value trend[b] | p-value[c] | p-value[c] |
| **Residence** | | | | | | |
| % Black | 0.28 | 0.96 | 1.00 | 1.00 | 0.28 | 0.46 |
| % White | 0.45 | 0.71 | 0.62 | 0.28 | 0.01 | 0.31 |
| % Hispanic | 0.47 | 0.86 | 0.57 | 0.37 | 0.01 | 0.17 |
| % Less than high school education | 0.03 | 0.84 | 0.20 | 0.48 | 0.00 | 0.11 |
| % Renter owner occupied | 0.40 | 0.80 | 0.98 | 0.72 | 0.00 | 0.08 |
| Median Household Income | 0.57 | 0.27 | 0.26 | 0.12 | 0.00 | 0.02 |
| Median Home Value | 0.51 | 0.16 | 0.31 | 0.21 | 0.01 | 0.28 |
| **Work** | | | | | | |
| % Black | 0.45 | 0.41 | 0.88 | 0.83 | 0.00 | 0.74 |
| % White | 0.37 | 0.36 | 0.42 | 0.46 | 0.40 | 0.78 |
| % Hispanic | 0.76 | 0.90 | 0.91 | 0.45 | 0.00 | 0.21 |
| % Less than high school education | 0.50 | 0.90 | 0.83 | 0.71 | 0.00 | 0.08 |

[a] *p-value was calculated using the Kruskal–Wallis test (non-normally distributed variables, as determined by a Shapiro-Wilk test)*
[b] *p for trend: p-value for trend test using the Jonckheere–Terpstra test.*
[c] *p-value was calculated using the Mann-Whitney U-test (non-normally distributed variables, as determined by a Shapiro-Wilk test)*
*Gray cells signify $p < 0.05$*
*Groups with significant differences using post-hoc Dunn tests:*

- *For marijuana facilities per km$^2$: quartiles 1 and 2, quartiles 2 and 4, and quartiles 3 and 4 for residents with formal education < high school*

***Table S4**: Results from the Mann-Whitney U test to compare the distribution of environmental justice related variables for census blockgroups containing complaints, versus those without, for every year between 2014-2022 at the census blockgroup level in Denver*

| | 2014 | 2015 | 2016 | 2017 | 2018 | 2019 | 2020 | 2021 | 2022 |
|---|---|---|---|---|---|---|---|---|---|
| | Complaints | Complaints | Complaints | Complaints | Complaints | Complaints | Complaints | Complaints | Complaints |
| | Complaints (yes/no) | Complaints (yes/no) | Complaints (yes/no) | Complaints (yes/no) | Complaints (yes/no) | Complaints (yes/no) | Complaints (yes/no) | Complaints (yes/no) | Complaints (yes/no) |
| | p-value[c] | p-value[c] | p-value[c] | p-value[c] | p-value[c] | p-value[c] | p-value[c] | p-value[c] | p-value[c] |
| **Residential** | | | | | | | | | |
| % Black | 0.92 | 0.88 | 0.89 | 0.49 | 0.59 | 0.77 | 0.87 | 0.88 | 0.14 |
| % White | 0.32 | 0.29 | 0.28 | 0.22 | 0.50 | 0.35 | 0.52 | 0.45 | 0.72 |
| % Hispanic | 0.49 | 0.46 | 0.09 | 0.11 | 0.35 | 0.14 | 0.79 | 0.36 | 0.91 |
| % Less than high school education | 0.57 | 0.88 | 0.34 | 0.27 | 0.83 | 0.82 | 0.82 | 0.71 | 0.90 |
| % Renter owner occupied | 0.06 | 0.12 | 0.08 | 0.40 | 0.09 | 0.38 | 0.14 | 0.90 | 0.80 |
| Median Household Income | 0.13 | 0.23 | 0.44 | 0.22 | 0.83 | 0.66 | 0.49 | 0.41 | 0.53 |
| Median Home Value | 0.92 | 0.56 | 0.59 | 0.56 | 0.32 | 0.54 | 0.32 | 0.18 | 0.77 |
| **Workplace** | | | | | | | | | |
| % Black | 0.10 | 0.14 | 0.81 | 0.44 | 0.05 | 0.65 | 0.24 | 0.80 | 0.40 |
| % White | 0.17 | 0.31 | 0.57 | 0.22 | 0.40 | 0.72 | 0.37 | 0.64 | 0.41 |
| % Hispanic | 0.46 | 0.87 | 0.13 | 0.05 | 0.75 | 1.00 | 0.34 | 0.38 | 0.69 |
| % Less than high school education | 0.82 | 0.79 | 0.14 | 0.67 | 0.90 | 0.98 | 0.22 | 0.81 | 0.30 |

*[a] p-value was calculated using the Kruskal–Wallis test (non-normally distributed variables, as determined by a Shapiro-Wilk test)*
*[b] p for trend: p-value for trend test using the Jonckheere–Terpstra test.*
*[c] p-value was calculated using the Mann-Whitney U-test (non-normally distributed variables, as determined by a Shapiro-Wilk test)*

***Table S5****: Results from statistical tests to compare the distribution of environmental justice related variables by exposure intensity based on different exposure metrics based on the number of complaints registered every year between 2014-2022 at the census blockgroup level in Denver*

| | | | **Residential** | | | | | | | **Work** | | | |
|---|---|---|---|---|---|---|---|---|---|---|---|---|---|
| | | | **% Black** | **% White** | **% Hispanic** | **% Less than high school education** | **% Renter owner occupied** | **Median Household Income** | **Median Home Value** | **% Black** | **% White** | **% Hispanic** | **% Less than high school education** |
| **2014** | **Per sq km** | **p-value[a]** | 0.62 | 0.13 | 0.22 | 0.03 | 0.00 | 0.65 | 0.97 | 0.09 | 0.21 | 0.00 | 0.00 |
| | | **p-value trend[b]** | 0.83 | 0.02 | 0.04 | 0.01 | 0.00 | 0.96 | 0.86 | 0.02 | 0.11 | 0.00 | 0.00 |
| | **Per population** | **p-value[a]** | 0.05 | 0.00 | 0.07 | 0.02 | 0.43 | 0.05 | 0.05 | 0.34 | 0.99 | 0.36 | 0.87 |
| | | **p-value trend[b]** | 0.02 | 0.00 | 0.05 | 0.01 | 0.63 | 0.56 | 0.53 | 0.18 | 0.87 | 0.75 | 0.82 |
| **2015** | **Per sq km** | **p-value[a]** | 0.74 | 0.07 | 0.05 | 0.12 | 0.02 | 0.28 | 0.05 | 0.41 | 0.53 | 0.05 | 0.06 |
| | | **p-value trend[b]** | 0.73 | 0.08 | 0.03 | 0.04 | 0.04 | 0.37 | 1.00 | 0.43 | 0.61 | 0.04 | 0.01 |
| | **Per population** | **p-value[a]** | 0.03 | 0.00 | 0.01 | 0.01 | 0.03 | 0.00 | 0.00 | 0.48 | 0.73 | 0.60 | 1.00 |
| | | **p-value trend[b]** | 0.30 | 0.32 | 0.68 | 0.26 | 0.88 | 0.46 | 0.97 | 0.09 | 0.67 | 0.27 | 1.00 |
| **2016** | **Per sq km** | **p-value[a]** | 0.26 | 0.78 | 0.55 | 0.76 | 0.10 | 0.97 | 0.89 | 0.07 | 0.36 | 0.14 | 0.07 |
| | | **p-value trend[b]** | 0.62 | 0.42 | 0.33 | 0.43 | 0.03 | 0.94 | 0.71 | 0.23 | 0.25 | 0.05 | 0.03 |
| | **Per population** | **p-value[a]** | 0.97 | 0.59 | 0.55 | 0.47 | 0.29 | 0.37 | 0.63 | 0.10 | 0.70 | 0.92 | 0.74 |
| | | **p-value trend[b]** | 0.86 | 0.66 | 0.45 | 0.22 | 0.49 | 0.43 | 0.27 | 0.01 | 0.23 | 0.73 | 0.96 |
| **2017** | **Per sq km** | **p-value[a]** | 0.46 | 0.29 | 0.32 | 0.15 | 0.16 | 0.06 | 0.97 | 0.10 | 0.08 | 0.47 | 0.22 |
| | | **p-value trend[b]** | 0.20 | 0.16 | 0.16 | 0.02 | 0.06 | 0.89 | 0.97 | 0.23 | 0.92 | 0.15 | 0.11 |
| | **Per population** | **p-value[a]** | 0.00 | 0.02 | 0.11 | 0.03 | 0.85 | 0.60 | 0.25 | 0.12 | 0.80 | 0.42 | 0.94 |
| | | **p-value trend**[b] | 0.01 | 0.15 | 0.11 | 0.02 | 0.82 | 0.45 | 0.21 | 0.03 | 0.81 | 0.25 | 0.92 |
| | **Per sq km** | **p-value**[a] | 0.43 | 0.13 | 0.17 | 0.42 | 0.04 | 0.86 | 0.46 | 0.39 | 0.34 | 0.00 | 0.01 |
| | | **p-value trend**[b] | 0.84 | 0.15 | 0.18 | 0.25 | 0.02 | 0.97 | 0.69 | 0.15 | 0.75 | 0.00 | 0.00 |
| | **Per population** | **p-value**[a] | 0.40 | 0.67 | 1.00 | 0.09 | 0.39 | 0.96 | 0.59 | 0.89 | 0.66 | 0.67 | 0.44 |
| **2018** | | **p-value trend**[b] | 0.12 | 0.52 | 0.89 | 0.04 | 0.96 | 0.87 | 0.52 | 0.70 | 0.58 | 0.63 | 0.28 |
| | **Per sq km** | **p-value**[a] | 0.68 | 0.02 | 0.05 | 0.00 | 0.17 | 0.13 | 0.09 | 0.04 | 0.21 | 0.00 | 0.03 |
| | | **p-value trend**[b] | 0.55 | 0.01 | 0.03 | 0.00 | 0.10 | 0.79 | 0.08 | 0.01 | 0.05 | 0.01 | 0.07 |
| | **Per population** | **p-value**[a] | 0.26 | 0.01 | 0.12 | 0.02 | 0.18 | 0.15 | 0.04 | 0.02 | 0.34 | 0.96 | 0.51 |
| **2019** | | **p-value trend**[b] | 0.06 | 0.00 | 0.02 | 0.00 | 0.67 | 0.14 | 0.01 | 0.00 | 0.06 | 1.00 | 0.47 |
| | **Per sq km** | **p-value**[a] | 0.12 | 0.13 | 0.30 | 0.05 | 0.02 | 0.02 | 0.05 | 0.50 | 0.83 | 0.09 | 0.14 |
| | | **p-value trend**[b] | 0.18 | 0.03 | 0.07 | 0.01 | 0.00 | 0.06 | 0.28 | 0.79 | 0.57 | 0.01 | 0.04 |
| | **Per population** | **p-value**[a] | 0.00 | 0.54 | 0.89 | 0.75 | 0.26 | 0.15 | 0.01 | 0.53 | 0.90 | 0.40 | 0.26 |
| **2020** | | **p-value trend**[b] | 0.07 | 0.59 | 0.62 | 0.55 | 0.54 | 0.55 | 0.73 | 0.18 | 0.69 | 0.48 | 0.58 |
| | **Per sq km** | **p-value**[a] | 0.76 | 0.05 | 0.12 | 0.10 | 0.46 | 0.66 | 0.11 | 0.20 | 0.96 | 0.23 | 0.04 |
| | | **p-value trend**[b] | 0.33 | 0.01 | 0.02 | 0.02 | 0.11 | 0.25 | 0.02 | 0.61 | 0.95 | 0.04 | 0.00 |
| | **Per population** | **p-value**[a] | 0.14 | 0.17 | 0.29 | 0.24 | 0.40 | 0.37 | 0.41 | 0.48 | 0.53 | 0.43 | 0.15 |
| **2021** | | **p-value trend**[b] | 0.05 | 0.14 | 0.55 | 0.18 | 0.53 | 0.22 | 0.18 | 0.19 | 0.95 | 0.09 | 0.20 |
| | **Per sq km** | **p-value**[a] | 0.92 | 0.99 | 0.97 | 0.75 | 0.03 | 0.33 | 0.13 | 0.90 | 0.80 | 0.84 | 0.46 |
| | | **p-value trend**[b] | 0.90 | 0.80 | 0.84 | 0.46 | 0.02 | 0.37 | 0.14 | 0.24 | 0.59 | 0.02 | 0.40 |
| | **Per population** | **p-value**[a] | 0.65 | 0.07 | 0.09 | 0.07 | 0.67 | 0.16 | 0.60 | 0.57 | 0.41 | 0.03 | 0.38 |
| **2022** | | **p-value trend**[b] | 0.24 | 0.12 | 0.07 | 0.29 | 0.46 | 0.52 | 0.55 | 0.05 | 0.92 | 0.17 | 0.54 |

*[a] p-value was calculated using the Kruskal–Wallis test (non-normally distributed variables, as determined by a Shapiro-Wilk test)*

*[b] p for trend: p-value for trend test using the Jonckheere–Terpstra test.*

*Gray cells signify $p < 0.05$*